\documentclass[aps,prb,reprint,superscriptaddress,nofootinbib,floatfix]{revtex4-2}

\usepackage{graphicx}
\usepackage{dcolumn}
\usepackage{bm}
\usepackage{mathrsfs}
\usepackage{amsthm,amssymb}
\usepackage{mathtools}
\usepackage{comment}
\usepackage[per-mode=fraction]{siunitx}
\usepackage[normalem]{ulem}
\usepackage{xcolor}
\usepackage{soul}
\usepackage{hyperref}
\hypersetup{
    colorlinks=true,
    linkcolor=blue,
    filecolor=magenta,
    urlcolor=cyan,
    citecolor=magenta,
}

\begin{document}

\title{Asymmetric Little--Parks Oscillations as a Probe
of Coupled Pairing Channels in Surface-Superconducting
\texorpdfstring{PtBi$_2$}{PtBi2}}

\author{Yuriy Yerin}
\affiliation{CNR-SPIN, via del Fosso del Cavaliere, 100, 00133 Roma, Italy}
\affiliation{Institute for Theoretical Solid State Physics, IFW Dresden, 01069 Dresden, Germany}

\author{Stefan-Ludwig Drechsler}
\affiliation{Institute for Theoretical Solid State Physics, IFW Dresden, 01069 Dresden, Germany}

\author{Carsten Timm}
\affiliation{Institute of Theoretical Physics, TUD Dresden University of Technology, 01062 Dresden, Germany}
\affiliation{W\"urzburg-Dresden Cluster of Excellence ctd.qmat, TUD Dresden University of Technology, 01062 Dresden, Germany}

\author{Jeroen van den Brink}
\affiliation{Institute for Theoretical Solid State Physics, IFW Dresden, 01069 Dresden, Germany}
\affiliation{Institute of Theoretical Physics, TUD Dresden University of Technology, 01062 Dresden, Germany}
\affiliation{W\"urzburg-Dresden Cluster of Excellence ctd.qmat, TUD Dresden University of Technology, 01062 Dresden, Germany}

\date{\today}

\begin{abstract}

Motivated by the recent discovery of unconventional surface superconductivity in PtBi$_{2}$, we theoretically study
its signatures in the Little–Parks oscillations of a narrow ring made from the same material. Using a Ginzburg--Landau model with three coupled pairing channels allowed by the trigonal crystal symmetry, we show that a fixed in-plane magnetic field can make the mean-field transition temperature asymmetric under reversal of the normal flux through the ring, while preserving periodicity in the superconducting flux quantum. The leading weak-field contribution to this asymmetry is cubic in the in-plane field strength. The flux-odd component respects the threefold crystal symmetry and changes sign when the in-plane field is rotated by $\pi/3$ at fixed magnitude. Within the homogeneous ring model, this asymmetry is absent for a single pairing channel and arises from the coupling between the three channels. Our calculations further show that increasing the in-plane magnetic field can enhance the superconducting nucleation temperature through mixing of pairing channels, illustrating magnetic control of multicomponent superconductivity in mesoscopic rings. Little--Parks interferometry with independent control of flux and field direction thus offers a mesoscopic probe of the pairing state, complementary to spectroscopic studies of the gap structure
\end{abstract}
\maketitle

% ============================================================
\section{Introduction}
\label{sec:intro}

The Little--Parks (LP) effect~\cite{LittleParks1962} is one of the clearest manifestations of macroscopic phase coherence in a multiply connected superconductor. In the conventional single-component Ginzburg--Landau (GL) description, fluxoid quantization makes the transition temperature periodic in the superconducting flux quantum $\Phi_0=hc/2e$, with a reciprocal response built from parabolic winding branches~\cite{Tinkham1996},
%\begin{equation}
$T_c(\phi)=T_c(-\phi),
$
%\qquad
with
$
\phi\equiv \Phi/\Phi_0.
$
%\end{equation}
Moreover, LP interferometry is a phase-sensitive probe of unconventional and multicomponent superconductivity. A familiar example is the $\pi$-shifted LP pattern, with extrema displaced from integer to half-integer flux, long discussed in connection with nontrivial pairing symmetry and multicomponent order~\cite{Geshkenbein1987,Kallin2016,Li2019Science,Almoalem2024}. In two-dimensional rings, vortices near the Berezinskii--Kosterlitz--Thouless (BKT) transition can produce half-quantum shifts even without unconventional pairing~\cite{XieNagaosa2026}, and half-quantum fluxoid features provide a further class of unconventional LP signatures~\cite{Yasui2017,Hua2023}. Unusual LP oscillations need not, however, imply a different flux period: multicomponent superconductors can preserve strict $\Phi_0$ periodicity while reshaping the branch structure, shifting avoided crossings, or changing the relative weight of winding sectors~\cite{Erin2008}.

Trigonal PtBi$_2$, in which superconductivity confined to the topological surface Fermi arcs has been observed~\cite{Kuibarov2024,Changdar2025} together with signatures of two-dimensional superconductivity~\cite{BKT_PtBi2}, provides a natural setting for unconventional LP oscillations. A multicomponent surface GL functional constrained by the point group $C_{3v}$, with superconducting sectors $A_1$, $A_2$, and $E$ and several symmetry-allowed magnetoelectric and Lifshitz invariants, has recently been proposed~\cite{Timm2025}. It builds on phenomenological and microscopic studies of Lifshitz invariants and magnetoelectric effects in noncentrosymmetric superconductors~\cite{MineevSamokhin1994,Mineev2008,Bychkov1984} and on the theory of polar superconductors~\cite{Edelstein1996}. Of particular interest are field-dependent gradient coupling terms that mix the one-component $A_1$ or $A_2$ order parameter with spatial variations of the two-component $E$ order parameter.
In a ring, these terms couple the order-parameter modes with
different winding numbers, describing different phase windings
around the circumference. Together with the other field-induced couplings, they modify the leading superconducting instability and can make the LP transition curve asymmetric under reversal of the normal flux at a fixed in-plane magnetic field.

The normal flux and the in-plane field $\mathbf B_\parallel = B\, (\cos\theta_B,\sin\theta_B,0)$ are independent control parameters,  where $B=|\mathbf B_\parallel|$ and $\theta_B$ is measured from the positive $x$ axis in the ring plane. Throughout this work $\phi\to-\phi$ is performed at fixed "in-plane" values of $B$ and $\theta_B$. A nonreciprocal Little--Parks response means
\begin{equation}
T_c(\phi,\mathbf B_\parallel)
\neq
T_c(-\phi,\mathbf B_\parallel).
\label{eq:intro_nonreciprocal}
\end{equation}
This asymmetry is compatible with magnetic time reversal,
which reverses both the normal flux and the in-plane
magnetic field:
\begin{equation}
T_c(\phi,\mathbf B_\parallel)
=
T_c(-\phi,-\mathbf B_\parallel).
\label{eq:intro_TR}
\end{equation}
Although motivated by PtBi$_2$, the proposed ring-based approach may also be applied to other layered noncentrosymmetric superconductors with suitable crystal symmetries and magnetoelectric couplings. Such a nonreciprocal response does not require a nodal superconducting gap.

We stress that the nonreciprocity under consideration
concerns the equilibrium transition temperature under normal-flux reversal, not a current reversal. It is also distinct from the superconducting diode effect~\cite{Ando2020,Nadeem2023} and from the nonreciprocal transport associated with LP oscillations in chiral superconducting nanotubes~\cite{Qin2017}. A superconducting diode effect has been observed in trigonal PtBi$_2$ for both in-plane and perpendicular field orientations~\cite{Jiang2025}. After an in-plane field is applied and removed, the effect persists at zero applied field, with its polarity controlled by the preceding field direction.

Other sources of LP asymmetry have been identified. Equilibrium LP phase shifts are predicted for rings partially covered by a ferromagnetic insulator, where the spatially nonuniform magnetoelectric coupling supplies a phase bias~\cite{Robinson2019}. Magnetic coupling to chiral order can reshape the LP boundary but without normal-state time-reversal-symmetry breaking this modulation is symmetric under field reversal~\cite{Xu2026}. Field misalignment provides a geometrical source of asymmetric patterns~\cite{Vekris2021}. 

The mechanism studied here is different: in a planar ring with spatially uniform GL coefficients, intercomponent winding coupling generate a flux-odd response that is cubic in the in-plane field at weak field and whose angular dependence is locked to the trigonal crystal axes. 

We find that the LP oscillations remain strictly $\Phi_0$-periodic but acquire a flux-odd component. This asymmetry arises from the combined action of field-induced couplings that mix the pairing channels and couple modes with different winding numbers. Its leading weak-field contribution is proportional to the cube of the in-plane field strength, $B^3$. The angular dependence reflects the threefold crystal symmetry: rotating the in-plane field by $\pi/3$ at fixed magnitude reverses the sign of the flux-odd component. In an ideal homogeneous ring, a single pairing channel yields a strictly reciprocal LP curve; consequently, the observed asymmetry indicates an interaction between the pairing channels. Finite flux can also raise the transition temperature above its zero-flux value at the same in-plane field. For the representative parameters studied here, this enhancement is generally smaller than the amplitude of the flux-odd component and varies strongly with the field and ring radius. The resulting signatures are compatible with the proposed $l=6$ $i$-wave pairing in PtBi$_2$~\cite{Changdar2025,Timm2025} without singling it out and complement probes of its nodal gap structure.

The paper is organized as follows: Section~\ref{sec:GL} summarizes the quadratic multicomponent GL functional, Sec.~\ref{sec:ring} derives its ring projection and winding-space structure, and Sec.~\ref{sec:Tc} formulates the generalized eigenvalue problem for $T_c$. Numerical results are presented in Sec.~\ref{sec:results}, followed by a discussion in Sec.~\ref{sec:discussion} and conclusions in Sec.~\ref{sec:conclusions}. Appendix~\ref{app:quadratic_invariants} discusses the choice of quadratic invariants and boundary contributions. Appendices~\ref{app:cartesian_projection} and \ref{app:analytic_origin} provide the numerical implementation and the analytical symmetry and perturbative analysis, respectively.

% ============================================================
\section{Ginzburg--Landau free energy}
\label{sec:GL}
% ============================================================

We start from the multicomponent Ginzburg--Landau functional
\begin{equation}
F[\psi_1,\psi_2,\boldsymbol{\psi}_E,\mathbf{B}]
= \int d^2r\;\mathcal{F}(\mathbf{r}),
\end{equation}
with gauge-covariant derivative
\begin{equation}
\mathbf D=\boldsymbol{\nabla}
-i\frac{q}{\hbar c}\mathbf A,
\qquad q=-2e,
\end{equation}
where $q$ is the Cooper-pair charge, $e>0$ is the elementary charge, and $\mathbf A$ is the vector potential. The two-component $E$-sector order parameter is
\begin{equation}
\boldsymbol{\psi}_E=(\psi_x,\psi_y).
\end{equation}
and the order parameters $\psi_1$ and $\psi_2$ transform as the $A_1$ and $A_2$ irreducible representations of $C_{3v}$, respectively. These order parameters label crystal-symmetry channels, not spin-singlet or spin-triplet sectors. In particular, the $A_2$ channel can contain both singlet and triplet contributions while producing the nodal $i$-wave surface gap discussed in Ref.~\cite{Changdar2025}.
Following Ref.~\cite{Timm2025}, we use an effective order parameter for each representation and do not resolve its microscopic singlet--triplet composition.

Near the superconducting transition, the normal state
\begin{equation}
\psi_1 = \psi_2 = 0, \qquad \boldsymbol{\psi}_E=\mathbf{0}
\end{equation}
loses stability when the lowest eigenvalue of the linearized GL operator passes through zero upon cooling. We denote the corresponding mean-field transition temperature by $T_c$.
In a finite, effectively one-dimensional ring, thermal amplitude and phase fluctuations broaden the superconducting onset, while phase slips can produce resistance below $T_c$. A possible BKT transition instead concerns vortex--antivortex unbinding in an effectively two-dimensional system and is not assumed in the present one-dimensional model. Consequently, the measured resistance need not be described by a transition curve that shifts rigidly with the calculated $T_c(\phi,\mathbf B_\parallel)$.

We parametrize the temperature dependence of the phenomenological coefficients as
\begin{equation}
\alpha_a(T)=a_a\bigl(T-T_{c,a}^{(0)}\bigr),
\quad
a_a>0,
\quad
a\in\{1,2,E\},
\end{equation}
and treat all remaining GL coefficients as temperature-independent constants. Here, $T_{c,a}^{(0)}$ is the transition temperature of the uncoupled, spatially uniform channel $a$, and $a_a$ is its coefficient. The actual LP transition temperature is obtained from the coupled problem and need not coincide with any $T_{c,a}^{(0)}$.

To isolate a concrete mechanism, we preserve the quadratic invariants listed below and set additional inter-sector cross-gradient terms and higher-order couplings to zero. The quadratic free-energy density is then given by \cite{Timm2025}
\begin{equation}
\begin{split}
\mathcal{F}^{(2)}={}&F_1+F_2+F_E+F^{(L)}_{1E}+F^{(L)}_{2E}\\
&\quad+F^{(B)}_{1E}+F^{(B)}_{2E}+F^{(L)}_1+F^{(L)}_2+F^{(L)}_{12}\\
&\quad+F^{(L)}_E+F^{(L)'}_{1E}+F^{(L)'}_{2E}.
\end{split}
\label{eq:F2_decomp}
\end{equation}
The individual terms in Eq.~\eqref{eq:F2_decomp}
are specified below. All phenomenological coefficients
are real. We introduce the following combinations
of derivatives of the $E$ components:
\begin{align}
\Sigma &\equiv D_x\psi_x+D_y\psi_y,
\label{eq:app_Sigma_def}\\
\Delta &\equiv D_x\psi_y-D_y\psi_x,
\label{eq:app_Delta_def}\\
C_1 &\equiv D_x\psi_y+D_y\psi_x,
\label{eq:app_C1_def}\\
C_2 &\equiv D_x\psi_x-D_y\psi_y.
\label{eq:app_C2_def}
\end{align}
Here, $\Sigma$ and $\Delta$ are the divergence- and curl-like combinations, while $C_1$ and $C_2$ describe the symmetric traceless gradient combinations.

The quadratic mass and gradient terms are
\begin{align}
F_1
&=\alpha_1(T)|\psi_1|^2
+\gamma_1\Bigl(|D_x\psi_1|^2+|D_y\psi_1|^2\Bigr),
\label{eq:app_F1}\\
F_2
&=\alpha_2(T)|\psi_2|^2
+\gamma_2\Bigl(|D_x\psi_2|^2+|D_y\psi_2|^2\Bigr),
\label{eq:app_F2}\\
F_E
&=\alpha_E(T)\Bigl(|\psi_x|^2+|\psi_y|^2\Bigr)
\nonumber\\
&\quad+\gamma_E|\Sigma|^2+\gamma'_E|\Delta|^2.
\label{eq:app_FE}
\end{align}
The additional symmetry-allowed gradient term
$\gamma''_E(|C_1|^2+|C_2|^2)$ is set to zero here.
Its relation to the effective tangential stiffnesses
is discussed in Appendix~\ref{app:quadratic_invariants}.

The zero-field mixed Lifshitz terms couple the scalar
channels to derivatives of the $E$ components:
\begin{align}
F^{(L)}_{1E}
&=\eta_{1E}\Bigl(\psi_1^*\Sigma+\Sigma^*\psi_1\Bigr),
\label{eq:app_F1EL}\\
F^{(L)}_{2E}
&=\eta_{2E}\Bigl(\psi_2^*\Delta+\Delta^*\psi_2\Bigr).
\label{eq:app_F2EL}
\end{align}

Writing the in-plane field as $\mathbf B_\parallel=(B_x,B_y)$, the field-linear couplings without derivatives are
\begin{align}
F^{(B)}_{1E}
&=i\zeta_{1E}\Bigl[
B_y\bigl(\psi_1^*\psi_x-\psi_x^*\psi_1\bigr)
\nonumber\\
&\qquad
-B_x\bigl(\psi_1^*\psi_y-\psi_y^*\psi_1\bigr)
\Bigr],
\label{eq:app_F1EB}\\
F^{(B)}_{2E}
&=i\zeta_{2E}\Bigl[
B_x\bigl(\psi_2^*\psi_x-\psi_x^*\psi_2\bigr)
\nonumber\\
&\qquad
+B_y\bigl(\psi_2^*\psi_y-\psi_y^*\psi_2\bigr)
\Bigr].
\label{eq:app_F2EB}
\end{align}

The field-dependent helical Lifshitz terms in the
scalar sectors are
\begin{align}
F^{(L)}_1
&=i\xi_1\Bigl[
B_y\psi_1^*D_x\psi_1
-B_x\psi_1^*D_y\psi_1
-\mathrm{c.c.}
\Bigr],
\label{eq:app_F1L}\\
F^{(L)}_2
&=i\xi_2\Bigl[
B_y\psi_2^*D_x\psi_2
-B_x\psi_2^*D_y\psi_2
-\mathrm{c.c.}
\Bigr],
\label{eq:app_F2L}\\
F^{(L)}_{12}
&=i\xi_{12}\Bigl[
B_x\psi_1^*D_x\psi_2
+B_y\psi_1^*D_y\psi_2
-\mathrm{c.c.}
\Bigr].
\label{eq:app_F12L}
\end{align}
Here and below, $\mathrm{c.c.}$ denotes the complex
conjugate of the preceding expression inside the
same brackets.

The corresponding terms within the $E$ channel are
\begin{equation}
\begin{split}
F^{(L)}_E
={}&i\xi'_E\Bigl[
\bigl(B_y\psi_x^*-B_x\psi_y^*\bigr)\Sigma
-\mathrm{c.c.}
\Bigr]\\
&+i\xi''_E\Bigl[
\bigl(B_y\psi_y^*+B_x\psi_x^*\bigr)\Delta
-\mathrm{c.c.}
\Bigr].
\end{split}
\label{eq:app_FEL}
\end{equation}

Finally, the tensorial Lifshitz terms couple the
scalar channels to the combinations $C_1$ and $C_2$:
\begin{align}
F^{(L)'}_{1E}
&=i\xi'_{1E}\Bigl[
B_y\psi_1^*C_1-B_x\psi_1^*C_2
-\mathrm{c.c.}
\Bigr],
\label{eq:app_F1ELprime}\\
F^{(L)'}_{2E}
&=i\xi'_{2E}\Bigl[
B_y\psi_2^*C_2+B_x\psi_2^*C_1
-\mathrm{c.c.}
\Bigr].
\label{eq:app_F2ELprime}
\end{align}
These terms involve the symmetric traceless gradient
combinations and generate couplings between winding
indices differing by two in the ring model of
Sec.~\ref{sec:ring}.

We use the expressions above as written.
Their relation to alternative forms obtained by
integration by parts, and the associated boundary
contributions, are discussed in
Appendix~\ref{app:quadratic_invariants}.

% ============================================================
\section{Ring-geometry reduction}
\label{sec:ring}
% ============================================================

We consider a narrow superconducting ring of radius $R$
in the $xy$ plane, parametrized by the polar angle $\theta$.
We treat it as an effectively one-dimensional system:
the order-parameter fields depend only on $\theta$, and
we retain only covariant derivatives along the circumference
in the quadratic GL terms. Their coefficients are
phenomenological parameters of the ring and need not
coincide with those of an extended superconducting surface.
A complete derivation from a finite-width annulus with
specified edge conditions is not attempted here.

For a normal (perpendicular) flux $\Phi$, the covariant derivative along
the circumference is
\begin{equation}
D_s=\frac{1}{R}\partial_\theta+i\frac{\phi}{R},
\qquad
\phi=\frac{\Phi}{\Phi_0},
\label{eq:Ds_ring}
\end{equation}
where $s=R\theta$ is the arc-length coordinate and $\Phi_0=hc/(2e)>0$. We distinguish the normal flux threading the ring from the local perpendicular magnetic field in the superconducting annulus. In the ideal flux-threading geometry, the latter vanishes, while the vector
potential has circulation $\oint\mathbf A\cdot d\mathbf l=\Phi$.
Thus, the flux enters through $D_s$, whereas explicit local-$B_z$ couplings are omitted. For a uniform perpendicular-field sweep, neglecting these couplings is an additional approximation.

In this one-dimensional model, the Cartesian derivatives
in the GL terms are replaced by
\[
D_x\longrightarrow-\sin\theta\,D_s,
\qquad
D_y\longrightarrow\cos\theta\,D_s,
\]
corresponding to differentiation along the local tangent
$\mathbf t=(-\sin\theta,\cos\theta)$.

For the $E$ doublet, it is convenient to use the components
along the local radial direction
$\mathbf n=(\cos\theta,\sin\theta)$ and the tangent.
The corresponding amplitudes $\psi_r$ and $\psi_t$ are
related to the Cartesian components by
\begin{align}
\psi_x&=\cos\theta\,\psi_r-\sin\theta\,\psi_t,
\nonumber\\
\psi_y&=\sin\theta\,\psi_r+\cos\theta\,\psi_t.
\label{eq:cart_local_basis}
\end{align}
When differentiating these expressions, the angular
dependence of the basis must be preserved.

The order-parameter components are single-valued,
$\psi_a(\theta+2\pi)=\psi_a(\theta)$, and can be expanded in 
a series of winding harmonics:
\begin{equation}
\psi_a(\theta)
=
\sum_{n\in\mathbb Z}\psi_{a,n}e^{in\theta},
\qquad
a\in\{1,2,r,t\}.
\label{eq:psi_angular_expansion}
\end{equation}
Here $\psi_{a,n}$ is the amplitude of the $n$th winding
harmonic of the component $a$ in the chosen ring basis.
Acting on a harmonic, the covariant derivative gives
\begin{equation}
D_s e^{in\theta}=ik_n e^{in\theta},
\qquad
k_n=\frac{n+\phi}{R}.
\label{eq:kn_def}
\end{equation}
Thus, the normal flux shifts the wave number entering the
gradient energy. We collect the amplitudes with the
same winding index into
\begin{equation}
\Psi_n=
(\psi_{1,n},\psi_{2,n},\psi_{r,n},\psi_{t,n})^T.
\label{eq:Psi_n_def}
\end{equation}

Although the phenomenological coefficients are constant
in the fixed crystal frame, the direction of the ring
tangent changes with $\theta$. This introduces angular
factors into the one-dimensional GL terms. For example,
the derivative combination in the scalar helical Lifshitz
terms becomes
\[
B_yD_x-B_xD_y
\longrightarrow
-B\cos(\theta-\theta_B)\,D_s.
\]
The cosine factor couples the winding harmonics with those
indices differing by one, since a multiplication of $e^{in\theta}$
by $e^{\pm i\theta}$ produces $e^{i(n\pm1)\theta}$.
This angular dependence reflects 
%the 
our ring geometry,
not a spatial variation of the material coefficients.

The tensorial Lifshitz terms involve the gradient combinations
$C_1=D_x\psi_y+D_y\psi_x$ and $C_2=D_x\psi_x-D_y\psi_y$.
Using the derivative replacements and the local-basis
transformation above gives
\[
\begin{aligned}
C_2+iC_1
&=
e^{2i\theta}
\left(iD_s-\frac{1}{R}\right)
(\psi_r+i\psi_t),\\
C_2-iC_1
&=
e^{-2i\theta}
\left(-iD_s-\frac{1}{R}\right)
(\psi_r-i\psi_t).
\end{aligned}
\]
Thus, the factors $e^{\pm2i\theta}$ make these terms
couple modes whose winding indices differ by two, $\Delta n=\pm2$. The curvature terms proportional to $1/R$ arise from differentiating the local basis. For example, the differential operator in the first expression acts on a winding harmonic as
\begin{align}
\left(iD_s-\frac{1}{R}\right)e^{in\theta}
&=
-\left(k_n+\frac{1}{R}\right)e^{in\theta}
\nonumber\\
&=-k_{n+1}e^{in\theta}.
\label{eq:kappa_real_main}
\end{align}
Thus, multiplication by $e^{2i\theta}$ changes the winding
index from $n$ to $n+2$, while the differential operator
supplies the coefficient $-k_{n+1}$. Further details of
the gradient reduction are given in
Appendix~\ref{app:cartesian_projection}.

The quadratic GL free energy in Eq.~\eqref{eq:F2_decomp}
gives the linearized GL equations $\widehat{\mathcal K}\Psi=0$, where $\Psi=(\psi_1,\psi_2,\psi_r,\psi_t)^T$.
In this component basis, the differential operator is
\begin{equation}
\widehat{\mathcal K}
=
\begin{pmatrix}
\widehat{\mathcal K}_{11} & \widehat{\mathcal K}_{12}
& \widehat{\mathcal K}_{1r} & \widehat{\mathcal K}_{1t}\\
\widehat{\mathcal K}_{12}^{\dagger} & \widehat{\mathcal K}_{22}
& \widehat{\mathcal K}_{2r} & \widehat{\mathcal K}_{2t}\\
\widehat{\mathcal K}_{1r}^{\dagger}
& \widehat{\mathcal K}_{2r}^{\dagger}
& \widehat{\mathcal K}_{rr} & \widehat{\mathcal K}_{rt}\\
\widehat{\mathcal K}_{1t}^{\dagger}
& \widehat{\mathcal K}_{2t}^{\dagger}
& \widehat{\mathcal K}_{rt}^{\dagger} & \widehat{\mathcal K}_{tt}
\end{pmatrix}.
\label{eq:GL_ring_operator_4x4}
\end{equation}
All entries are scalar differential operators. The diagonal
entries, with $a=1,2$, are
\begin{align}
\widehat{\mathcal K}_{aa}
={}&\alpha_a(T)-\gamma_aD_s^2
\nonumber\\
&-i\xi_a B\left[
2\cos(\theta-\theta_B)D_s
-\frac{\sin(\theta-\theta_B)}{R}
\right],
\label{eq:GL_4x4_scalar_diagonals}\\
\widehat{\mathcal K}_{rr}
={}&\alpha_E(T)-\gamma'_E D_s^2+\frac{\gamma_E}{R^2}
\nonumber\\
&-i\xi''_E B\left[
2\cos(\theta-\theta_B)D_s
-\frac{\sin(\theta-\theta_B)}{R}
\right],
\label{eq:GL_4x4_radial_diagonal}\\
\widehat{\mathcal K}_{tt}
={}&\alpha_E(T)-\gamma_E D_s^2+\frac{\gamma'_E}{R^2}
\nonumber\\
&-i\xi'_E B\left[
2\cos(\theta-\theta_B)D_s
-\frac{\sin(\theta-\theta_B)}{R}
\right].
\label{eq:GL_4x4_tangential_diagonal}
\end{align}

The couplings within the scalar and $E$ sectors are
\begin{align}
\widehat{\mathcal K}_{12}
={}&-i\xi_{12}B\sin(\theta-\theta_B)D_s,
\label{eq:GL_4x4_scalar_mixing}\\
\widehat{\mathcal K}_{rt}
={}&\frac{\gamma_E+\gamma'_E}{R}D_s
\nonumber\\
&-iB(\xi'_E-\xi''_E)\sin(\theta-\theta_B)D_s
\nonumber\\
&+\frac{iB(\xi'_E+2\xi''_E)}{R}\cos(\theta-\theta_B).
\label{eq:GL_4x4_rt_mixing}
\end{align}

The scalar--doublet couplings are
\begin{align}
\widehat{\mathcal K}_{1r}
={}&\frac{\eta_{1E}}{R}
-i\zeta_{1E}B\sin(\theta-\theta_B)
\nonumber\\
&+i\xi'_{1E}B\left[
\sin(2\theta+\theta_B)D_s
+\frac{\cos(2\theta+\theta_B)}{R}
\right],
\label{eq:GL_4x4_1r}\\
\widehat{\mathcal K}_{1t}
={}&\eta_{1E}D_s
-i\zeta_{1E}B\cos(\theta-\theta_B)
\nonumber\\
&+i\xi'_{1E}B\left[
\cos(2\theta+\theta_B)D_s
-\frac{\sin(2\theta+\theta_B)}{R}
\right],
\label{eq:GL_4x4_1t}\\
\widehat{\mathcal K}_{2r}
={}&-\eta_{2E}D_s
+i\zeta_{2E}B\cos(\theta-\theta_B)
\nonumber\\
&+i\xi'_{2E}B\left[
\cos(2\theta+\theta_B)D_s
-\frac{\sin(2\theta+\theta_B)}{R}
\right],
\label{eq:GL_4x4_2r}\\
\widehat{\mathcal K}_{2t}
={}&\frac{\eta_{2E}}{R}
-i\zeta_{2E}B\sin(\theta-\theta_B)
\nonumber\\
&-i\xi'_{2E}B\left[
\sin(2\theta+\theta_B)D_s
+\frac{\cos(2\theta+\theta_B)}{R}
\right].
\label{eq:GL_4x4_2t}
\end{align}

All derivatives act on the order-parameter components
to their right. The dagger denotes the operator adjoint
with respect to the inner product on the ring,
\[
\langle u,v\rangle
=\frac{1}{2\pi}\int_0^{2\pi}
u^\dagger(\theta)v(\theta)\,d\theta.
\]
The remaining matrix entries follow by taking the
corresponding operator adjoints.

The factors $\cos(\theta-\theta_B)$ and $\sin(\theta-\theta_B)$ couple winding indices differing by one,
whereas $\cos(2\theta+\theta_B)$ and $\sin(2\theta+\theta_B)$ couple terms with
indices differing by two. The remaining, angle-independent terms preserve the winding index.

Substituting the winding expansion Eq.
\eqref{eq:psi_angular_expansion} into the quadratic free
energy and integrating over $\theta$ gives the GL matrix
$\mathbb K$ in the winding basis. Its entries are
\begin{equation}
\begin{aligned}
\bigl[\mathbb K(T,\phi,\mathbf B_\parallel)\bigr]_{\mu n,\nu m}
={}&\frac{1}{2\pi}\int_0^{2\pi}d\theta\,
e^{-in\theta}\\
&\times
\left[
\widehat{\mathcal K}_{\mu\nu}
(T,\theta;\phi,\mathbf B_\parallel)e^{im\theta}
\right].
\end{aligned}
\label{eq:fourier_projection_kernel}
\end{equation}
Here $\mu,\nu\in\{1,2,r,t\}$ label the order-parameter
components. The differential operator acts on
$e^{im\theta}$ before the integration, which extracts
the $n^{\mbox{th}}$ Fourier coefficient. Each matrix element
therefore describes the coupling between the amplitudes
$\psi_{\nu,m}$ and $\psi_{\mu,n}$.

The angular factors in the operator above determine which
winding modes are coupled. We define
$\Delta n=n-m$, where $m$ and $n$ are the incoming and
outgoing winding indices, respectively. For the quadratic
GL terms retained here, only
$\Delta n=0,\pm1,\pm2$ occur. The quadratic free energy takes the form
\begin{equation}
F^{(2)}
=
wL\sum_{n,m}
\Psi_n^\dagger
\mathbb K_{nm}(T,\phi,\mathbf B_\parallel)
\Psi_m ,
\label{eq:F2_projected}
\end{equation}
where $w$ is the ring width and $L=2\pi R$ denotes the ring
%its
circumference. The overall factor $wL$ does not affect  $T_c$ we are looking for. The angle-independent terms contribute to the
$\Delta n=0$ blocks. The factors $\cos(\theta-\theta_B)$ and $\sin(\theta-\theta_B)$ in the field-dependent couplings generate
$\Delta n=\pm1$ blocks, whereas the second harmonics
$\cos(2\theta+\theta_B)$ and $\sin(2\theta+\theta_B)$ in the tensorial Lifshitz couplings generate $\Delta n=\pm2$ blocks.
The latter provide the additional winding-mode mixing
that is absent in the conventional scalar LP problem.

The resulting GL matrix satisfies
\begin{align}
\mathbb K(T,\phi,\mathbf B_\parallel)
&=
\mathbb K^\dagger(T,\phi,\mathbf B_\parallel),
\label{eq:kernel_hermiticity}\\
\mathbb K(T,\phi,\mathbf B_\parallel)
&=
P_n\,
\mathbb K^*(T,-\phi,-\mathbf B_\parallel)
P_n^\dagger ,
\label{eq:kernel_TR}
\end{align}
where $P_n$ reverses the winding index,
$n\to-n$, and acts trivially in the
$(1,2,r,t)$ component space.
Eq.~\eqref{eq:kernel_TR} implies the magnetic
time-reversal relation for the transition temperature,
Eq.~\eqref{eq:intro_TR}.

% ============================================================
\section{Eigenvalue problem for \texorpdfstring{$T_c(\Phi)$}{Tc(Phi)}}
\label{sec:Tc}
% ============================================================

Truncating the winding expansion to $|n|\leq N$ and stacking the sector vectors into $\mathcal V_N=(\Psi_{-N},\ldots,\Psi_N)^T$, the GL free energy becomes
\begin{equation}
F^{(2)}=wL\,\mathcal V_N^\dagger
\mathbb K_N(T,\phi,\mathbf B_\parallel)\mathcal V_N.
\end{equation}
For the given functional, the matrix $\mathbb K_N$ is block-pen\-ta\-di\-a\-go\-nal because the projection produces only $\Delta n=0,\pm1,\pm2$ matrix elements.  We write
\begin{equation}
\mathbb K_N(T,\phi,\mathbf B_\parallel)
=
\mathbb K_N^{(0)}(\phi,\mathbf B_\parallel)+T\,\mathbb S_N,
\label{eq:K_decomp}
\end{equation}
where $\mathbb S_N=\mathbf 1_{2N+1}\otimes\mathrm{diag}(a_1,a_2,a_E,a_E)$ is assembled from coefficients $a_a$ introduced in Sec.~\ref{sec:GL} and the superscript $(0)$ denotes the temperature-independent part of the GL matrix (it does not denote zero magnetic field).

The normal state becomes unstable when the lowest eigenvalue of $\mathbb K_N$ crosses zero.  Equivalently, the transition temperature is the largest generalized eigenvalue of
\begin{equation}
-\mathbb K_N^{(0)}(\phi,\mathbf B_\parallel)\,u
=
T\,\mathbb S_N u.
\label{eq:gen_eig}
\end{equation}
Thus, the critical temperature is given by
\begin{equation}
T_c(\phi,\mathbf B_\parallel)=
\max_j T_j(\phi,\mathbf B_\parallel),
\label{eq:Tc_max_eigenvalue}
\end{equation}
where $T_j$ are the real finite generalized eigenvalues of Eq.~\eqref{eq:gen_eig}.  The maximization over all branches is responsible for the conventional LP behavior.  Increasing the flux by one flux quantum can be compensated by decreasing every winding number by one, since $k_{n-1}(\phi+1)=k_n(\phi)$. This relabeling leaves the GL eigenvalue spectrum unchanged, so $T_c$ is exactly $\Phi_0$-periodic.
In the numerical calculations, we keep sufficiently many
winding modes to reproduce this periodicity to numerical
accuracy.

\begin{table*}[t]
\caption{Representative dimensionless GL coefficients used in the numerical figures. They are illustrative parameter choices, not fitted material constants due to the missing experimental realization of the proposed rings, so far. The geometric and field-control parameters $R$, $B$, and $\theta_B$ are specified separately in each figure caption.}
\label{tab:gl_params}
\begin{ruledtabular}
\begin{tabular}{lll}
Sector / coupling & Symbol & Value \\
\colrule
Bare critical temperatures & $T_{c,1}^{(0)},\ T_{c,2}^{(0)},\ T_{c,E}^{(0)}$ & $0.90,\ 1,\ 0.75$ \\
Mass prefactors & $a_1,\ a_2,\ a_E$ & $1,\ 1,\ 1$ \\
Gradient stiffnesses & $\gamma_1,\ \gamma_2,\ \gamma_E,\ \gamma'_E,\ \gamma''_E$ & $1,\ 0.9,\ 0.6,\ 0.45,\ 0$ \\
Zero-field mixing terms & $\eta_{1E},\ \eta_{2E}$ & $0,\ 0$ \\
Field-linear non-derivative terms & $\zeta_{1E},\ \zeta_{2E}$ & $-0.18,\ 0.15$ \\
Scalar-sector helical Lifshitz terms & $\xi_1,\ \xi_2,\ \xi_{12}$ & $-0.14,\ 0.08,\ -0.1$ \\
$E$-sector Lifshitz terms & $\xi'_E,\ \xi''_E$ & $-0.14,\ 0.12$ \\
Tensorial Lifshitz terms & $\xi'_{1E},\ \xi'_{2E}$ & $-0.18,\ -0.1$ \\
\end{tabular}
\end{ruledtabular}
\end{table*}

To describe the LP effect, we introduce three physical characteristics.  The finite-flux enhancement is
\begin{equation}
E(B,\theta_B)=
\max_\phi\bigl[T_c(\phi;B,\theta_B)-T_c(0;B,\theta_B)\bigr],
\label{eq:E_def}
\end{equation}
the nonreciprocal odd-in-flux amplitude is
\begin{equation}
A_{\rm nr}(B,\theta_B)=
\frac{1}{2}\max_\phi
\bigl|T_c(\phi;B,\theta_B)-T_c(-\phi;B,\theta_B)\bigr|,
\label{eq:Anr_def}
\end{equation}
and, at a fixed reference flux $\phi_0$, a signed odd response is
\begin{equation}
O_{\phi_0}(B,\theta_B)=
\frac{T_c(\phi_0;B,\theta_B)-T_c(-\phi_0;B,\theta_B)}{2}.
\label{eq:Ophi_def}
\end{equation}
Here, $\phi_0$ is a representative value of the reduced flux $\phi=\Phi/\Phi_0$, fixed while $B$ and $\theta_B$ are varied. It is chosen in the interval $0<\phi_0 < 1/2$ since the odd combination vanishes both at $\phi=0$ and, by $\Phi_0$ periodicity, at $\phi = 1/2$. Here we use $\phi_0 = 1/4$.  Unlike $A_{\rm nr}$, which takes the magnitude maximized over
the flux, $O_{\phi_0}$ preserves the sign and is useful for displaying the sign-changing angular structure of the nonreciprocal response.  All maximizations over
the flux (for $A_{\rm nr}$ and $E$) are performed over one period, i.e. $\phi\in[-1/2,1/2]$.

\begin{figure*}
\centering
\includegraphics[width=0.49\textwidth]{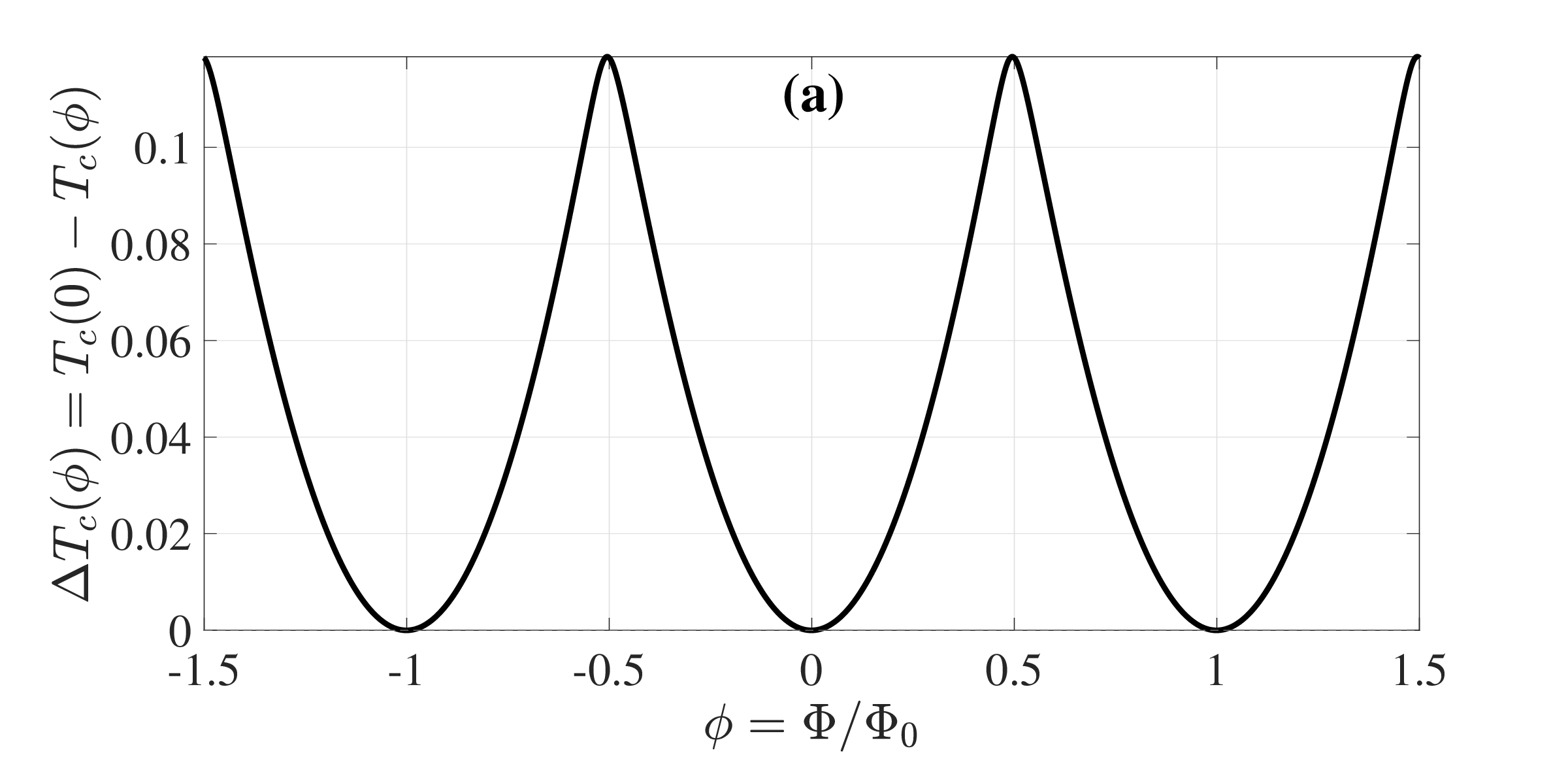}\hfill
\includegraphics[width=0.49\textwidth]{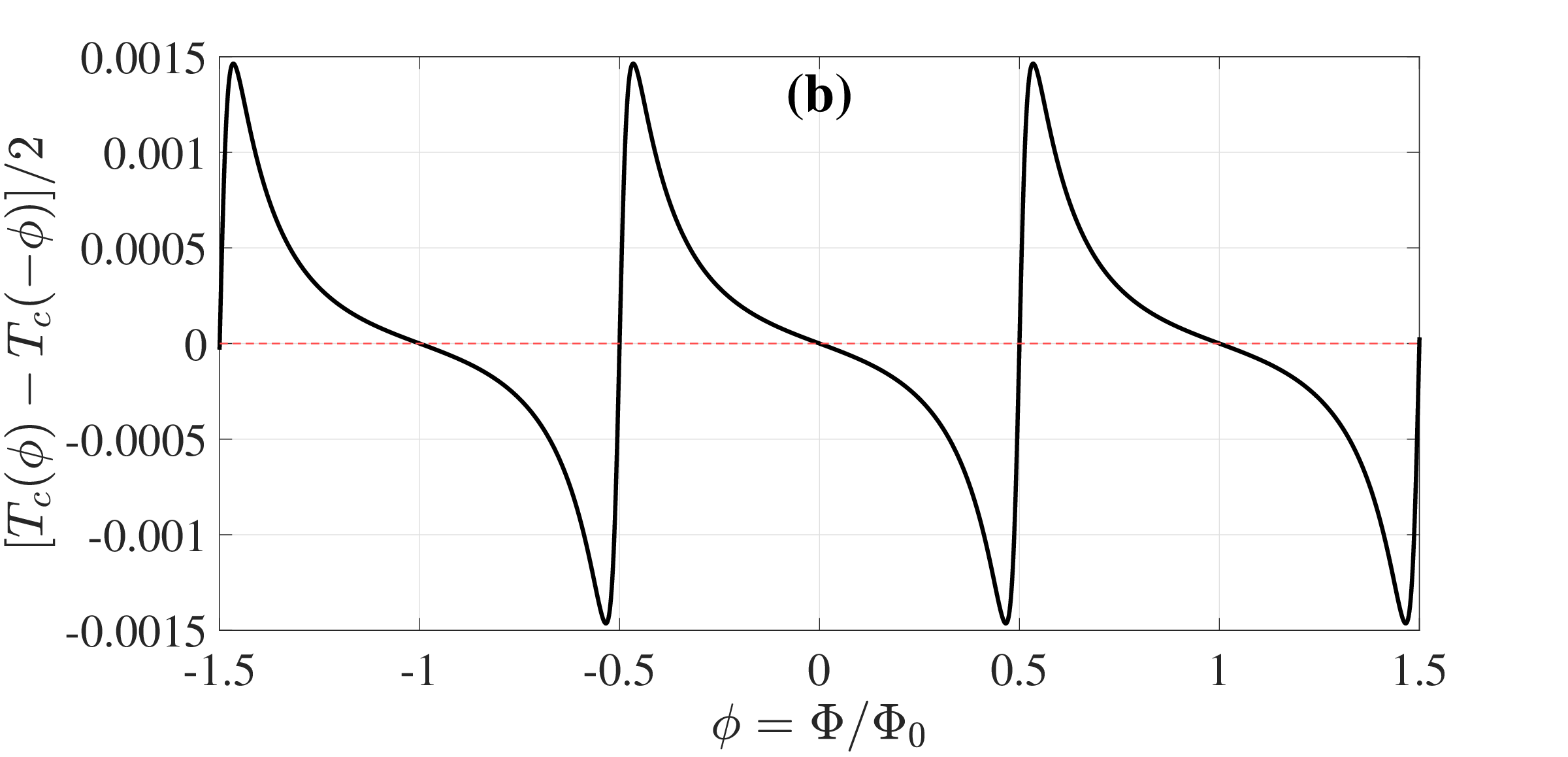}
\includegraphics[width=0.49\textwidth]{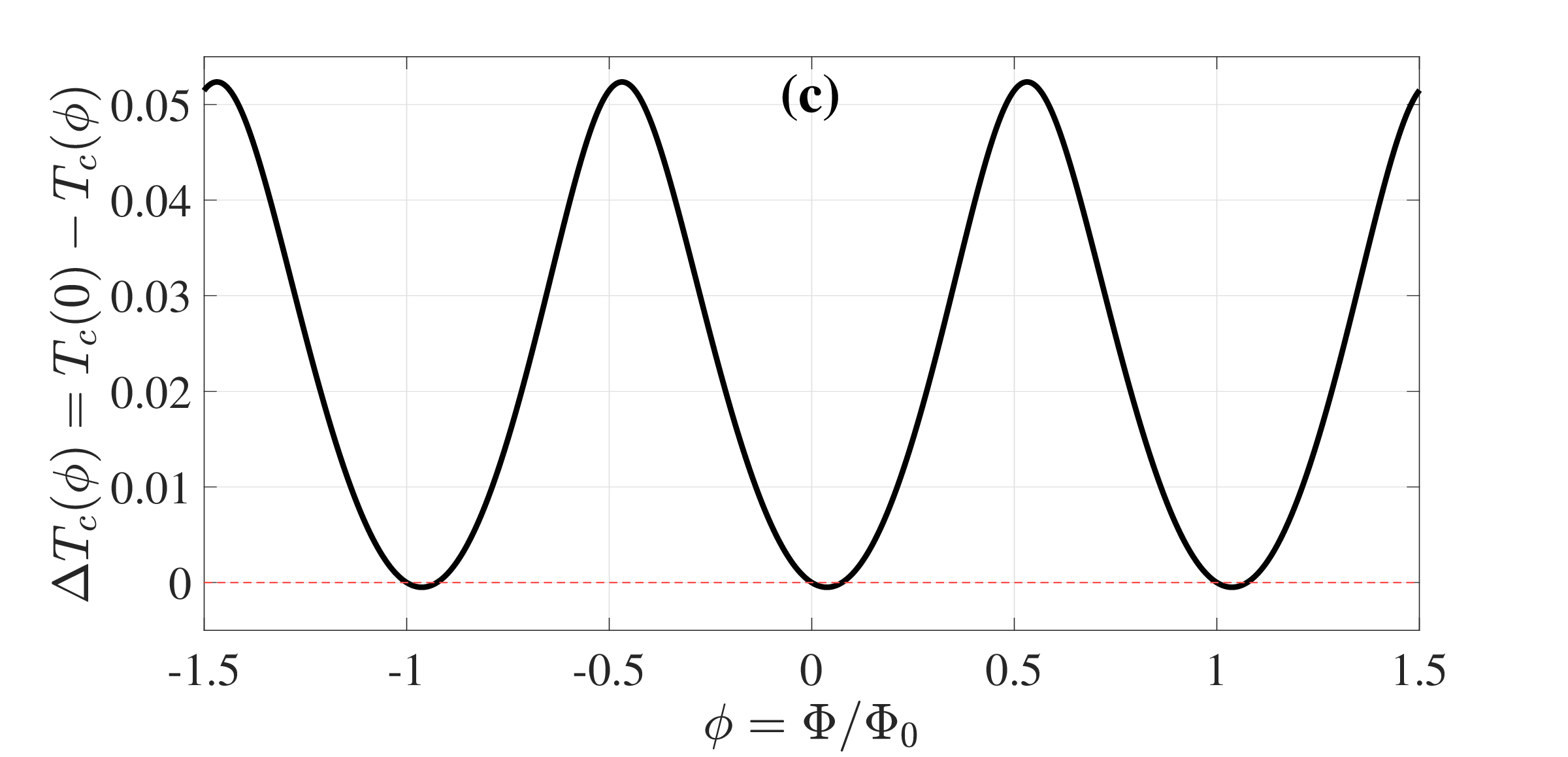}\hfill
\includegraphics[width=0.49\textwidth]{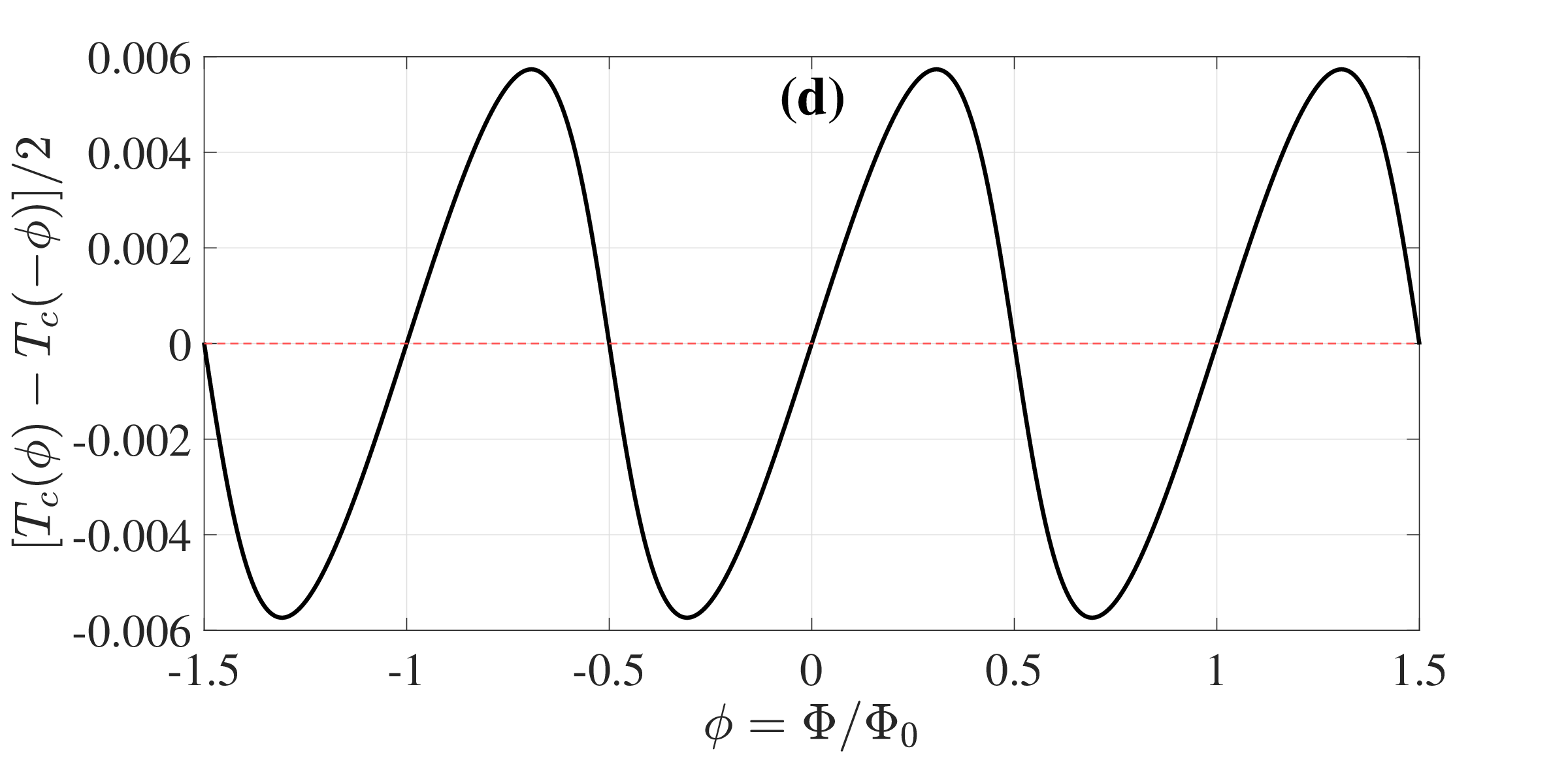}
\caption{(a), (c) LP shift $\Delta T_c(\phi)=T_c(0)-T_c(\phi)$ and (b), (d) flux-odd component $[T_c(\phi)-T_c(-\phi)]/2$ of $T_c$ at $R=1.25$ and $\theta_B=\pi/6$ for (a), (b) $B=0.75$ and (c), (d) $B=2$. The red dashed horizontal lines in panels (b)--(d) mark the zero level of the respective plotted quantities.}
\label{fig:linecut_cartesian}
\end{figure*}

The angular dependence of the flux-odd response is
constrained by the three mirror planes of $C_{3v}$.
Their intersections with the ring plane define mirror
lines at angles $\theta_m+k\pi/3$, where $\theta_m=\pi/6$ in our coordinates and $k=0,1,2$. When the in-plane field is perpendicular to a mirror line ($\theta_B=k\pi/3$), the reflection reverses the normal flux but leaves the in-plane field unchanged. Consequently,
$T_c(\phi,\mathbf B_\parallel)=T_c(-\phi,\mathbf B_\parallel)$, and both $A_{\rm nr}$ and $O_{\phi_0}$ vanish.
For a field parallel to a mirror line, reflection
reverses both magnetic controls and does not require
reciprocity at fixed in-plane field. In our calculations, we choose $\theta_B=\pi/6$, parallel to a mirror line.

% ============================================================
\section{Numerical results}
\label{sec:results}
% ============================================================

All numerical results in this section are obtained from the directly projected GL matrix of Sec.~\ref{sec:ring}. The truncation, quadrature, grids, and convergence checks are given in Appendix~\ref{app:cartesian_projection}.

Throughout, dimensionless units are used: temperatures are measured in units of the largest bare critical temperature, $T^{(0)}_{c,2}$, and lengths in units of
\begin{equation}
\xi_{\rm ref}=\sqrt{\gamma_1/(a_2T^{(0)}_{c,2})} ,
\end{equation}
with $\gamma_1$ and $a_2$ taken from Table~\ref{tab:gl_params}, where both equal unity. The in-plane field is measured in units of a reference scale $B_{\rm ref}$ that is absorbed into the field-linear coefficients. The dimensionless GL coefficients used for all figures in this section are collected in Table~\ref{tab:gl_params}. They are identical for every figure.

In the surface-pairing picture of
Refs.~\cite{Changdar2025,Timm2025}, the $A_2$ channel
corresponds to the proposed nodal $i$-wave gap profile.
The parameter hierarchy in Table~\ref{tab:gl_params}
selects an $A_2$ instability at zero field and zero flux.
Thus, for the parameters used here, the weak-field
nonreciprocal response originates from an $A_2$ ($i$-wave)
parent state coupled to the subdominant $A_1$ and $E$
pairing channels. However, neither the cubic dependence
on the in-plane field strength nor the trigonal angular
pattern is unique to $A_2$ pairing. Both are symmetry
allowed for other parent states. At higher fields, the
leading instability can also change its dominant
pairing-channel character.

We first plot the LP shift
\begin{equation}
\Delta T_c(\phi)=T_c(0)-T_c(\phi),
\label{eq:DeltaTc_def}
\end{equation}
defined in such a way that positive $\Delta T_c$ denotes the usual flux-induced suppression of $T_c$, whereas negative values indicate a finite-flux enhancement.

Figure~\ref{fig:linecut_cartesian} illustrates the main effect at two in-plane magnetic fields.  The LP oscillations remain periodic in $\Phi_0$ and, for the representative parameter set, acquire a finite odd-in-flux component.  At $B=0.75$ [Figs.~\ref{fig:linecut_cartesian}(a) and \ref{fig:linecut_cartesian}(b)], it becomes largest close to
a half-integer flux. At $B=2$ [Figs.~\ref{fig:linecut_cartesian}(c) and \ref{fig:linecut_cartesian}(d)], the odd component has the opposite sign and is about four times larger. 

This sign change accompanies a rearrangement of the leading instability, that is, a change in its component character. Writing the leading eigenvector of Eq.~\eqref{eq:gen_eig} as $u=(\Psi_{-N},\ldots,\Psi_N)^T$, with $\Psi_n$ given by Eq.~\eqref{eq:Psi_n_def}, we define the component weights
\begin{equation}
w_a=\frac{a_a\sum_n|\psi_{a,n}|^2}{u^\dagger\,\mathbb S_N\,u},
\qquad
a\in\{1,2,r,t\},
\label{eq:mode_weights}
\end{equation}
where $a_r=a_t=a_E$ so that $w_1+w_2+w_E=1$ with $w_E=w_r+w_t$. The $E$ weight $w_E$ does not depend on the in-plane basis chosen for the doublet. 

\begin{figure}[tb]
\centering
\includegraphics[width=\columnwidth]{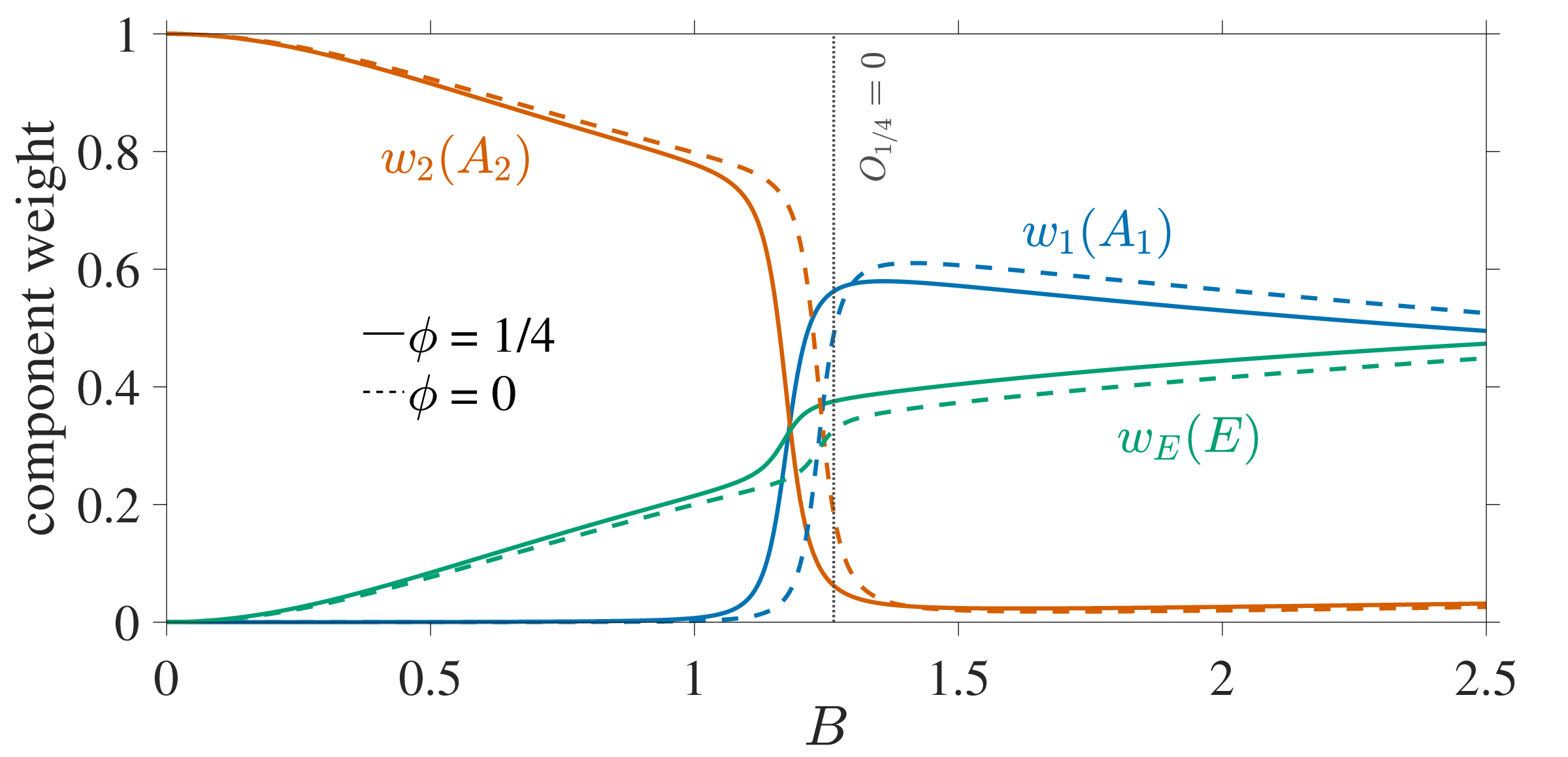}
\caption{Component content of the leading instability. Weights $w_1$, $w_2$, and $w_E=w_r+w_t$, Eq.~\eqref{eq:mode_weights}, vs.\ in-plane field at $\phi=1/4$ (solid) and $\phi=0$ (dashed), for $R=1.25$ and $\theta_B=\pi/6$. The $A_2$ weight decreases from unity to a few percent across the mode-reorganization region, while the $A_1$ and $E$ components take over. The dotted line marks the zero of the $O_{1/4}$ response at $B\simeq1.26$ (see Fig. \ref{fig:Anr_signed_maps}, center). Phenomenological coefficients are given in Table~\ref{tab:gl_params}.}
\label{fig:mode_content}
\end{figure}

\begin{figure*}[t]
\centering
\includegraphics[width=0.333\textwidth]{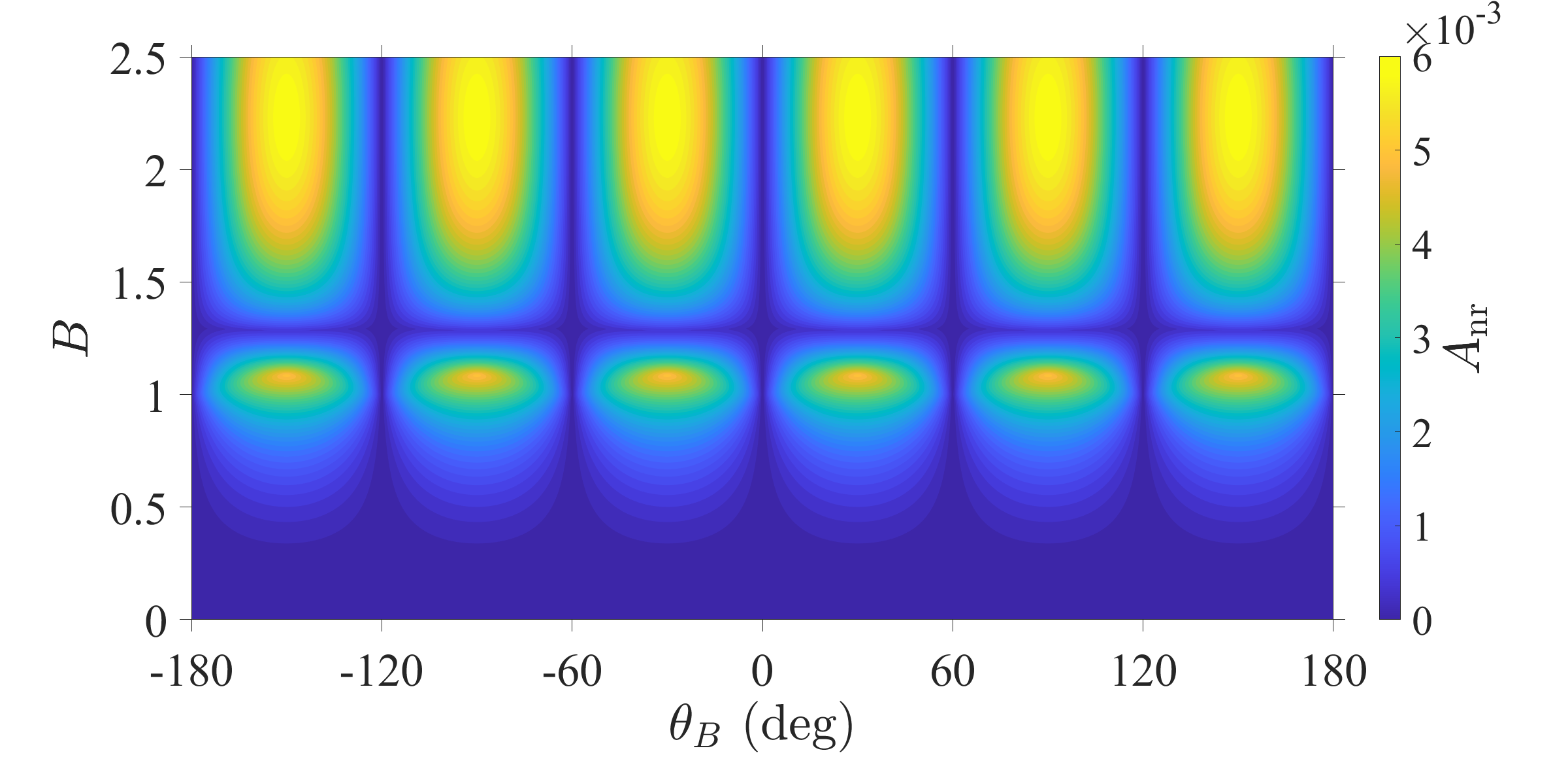}\hfill
\includegraphics[width=0.333\textwidth]{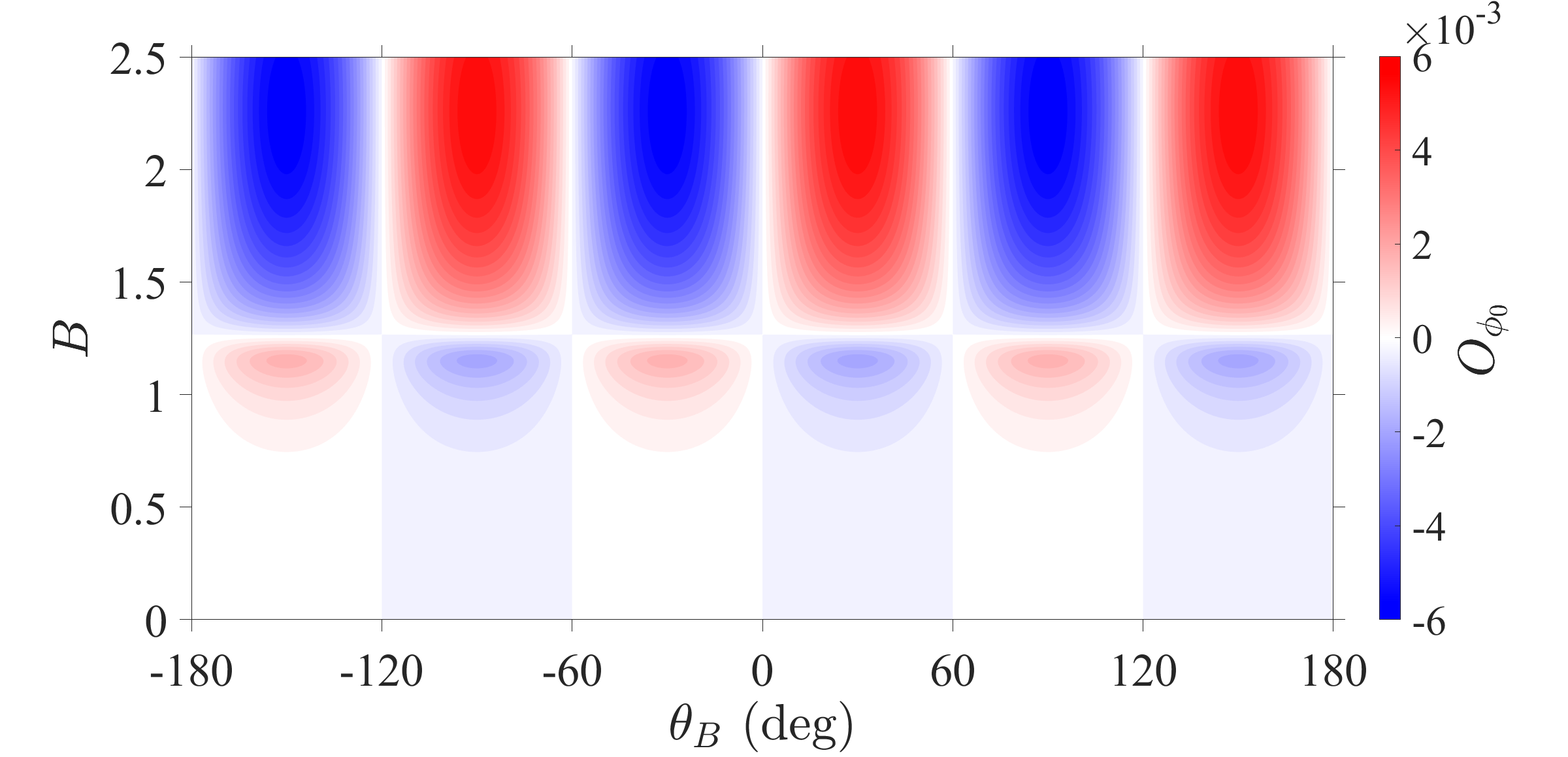}\hfill
\includegraphics[width=0.333\textwidth]{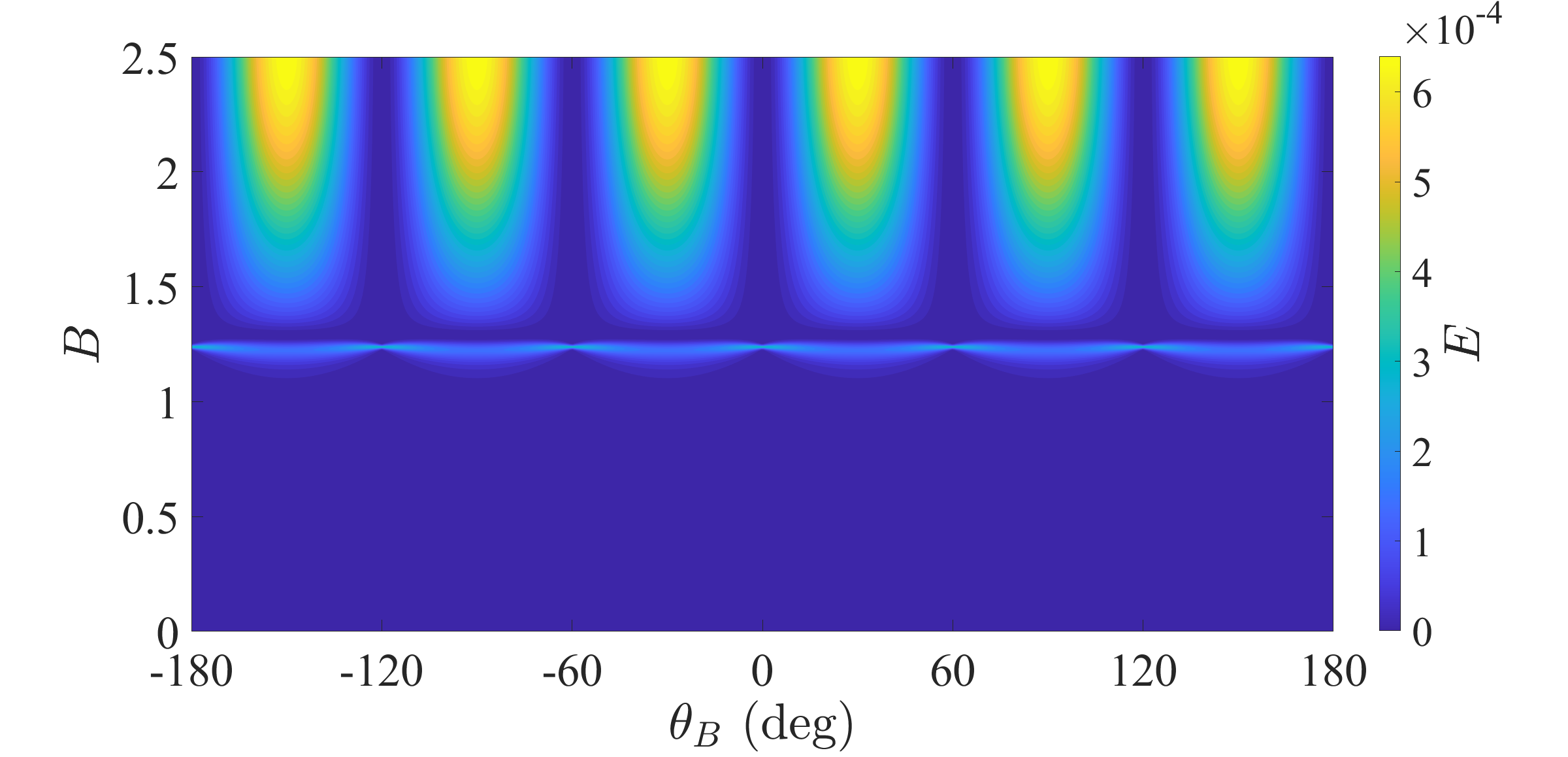}
\caption{Field-angle dependence of the nonreciprocal LP response. Left: absolute odd-in-flux amplitude $A_{\rm nr}(B,\theta_B)$. Center: signed odd response $O_{\phi_0}(B,\theta_B)$ at $\phi_0=1/4$, displaying alternating angular lobes and a polarity reversal in the mode-reorganization region near $B\simeq1.26$. Right: finite-flux enhancement $E(B,\theta_B)=\max_\phi[T_c(\phi)-T_c(0)]$, which is weak below the mode-reorganization region, shows a narrow ridge within it, and grows beyond it. The radius is fixed to $R=1.25$, the other coefficients are given in Table~\ref{tab:gl_params}.}
\label{fig:Anr_signed_maps}
\end{figure*}

Figure~\ref{fig:mode_content} shows these weights at
$\theta_B=\pi/6$. At weak fields, the leading instability
is dominated by the $A_2$ channel, with a field-induced
admixture of the $E$ components. As the in-plane field
increases, a mode dominated by the $A_1$ and $E$ components
acquires a higher transition temperature and becomes the
leading instability between $B\simeq1.1$ and $B\simeq1.3$.
At both $\phi=0$ and $\phi=1/4$, the two leading eigenvalues
approach each other without crossing, resulting in an
avoided crossing. The field at which $w_2=1/2$ depends weakly on flux. The zero of $O_{1/4}$ at $B\simeq1.26$, marked by the
dotted line in Fig.~\ref{fig:mode_content}, lies near
the upper end of this crossover.
At the mirror-preserving orientations $\theta_B=k\pi/3$,
the situation is different: at zero flux, the two leading
modes have opposite mirror parities and cross at
$B\simeq1.24$. A finite flux breaks this mirror symmetry
and turns the crossing into an avoided one.

For the coefficients in Table~\ref{tab:gl_params},
the odd-in-flux response grows as $B^3$ at sufficiently
weak in-plane fields and originates from an
$A_2$-dominated instability. At higher fields, the larger
odd amplitudes are associated with the field-reorganized
mode dominated by the $A_1$ and $E$ components.
The two field values used in Fig.~\ref{fig:linecut_cartesian}
illustrate this evolution. At $R=1.25$ and
$\theta_B=\pi/6$, the $A_2$ weight varies between
$0.72$ and $0.86$ over the flux period at $B=0.75$,
whereas it remains below $0.08$ at $B=2$.
Thus, the larger nonreciprocal response at $B=2$
characterizes the field-reorganized instability
rather than the nearly pure $A_2$ parent state.

We now examine the field and angular dependence of the LP
response in Fig.~\ref{fig:Anr_signed_maps}. The three
panels show the absolute odd amplitude
$A_{\rm nr}(B,\theta_B)$, Eq.~\eqref{eq:Anr_def},
the signed response $O_{\phi_0}(B,\theta_B)$,
Eq.~\eqref{eq:Ophi_def}, and the finite-flux
enhancement $E(B,\theta_B)$, Eq.~\eqref{eq:E_def},
respectively.

The map of $A_{\rm nr}$ is one of the central numerical results. It exhibits clear angular lobes
and an overall growth with increasing in-plane field,
with a suppression in the mode-reorganization region.
The signed response $O_{\phi_0}$ reverses sign near
$B\simeq1.26$, consistent with the change in the
leading instability shown in Fig.~\ref{fig:mode_content}.
The finite-flux enhancement $E$ remains small at lower
fields and exhibits a narrow ridge near $B\simeq1.22$--$1.24$, associated with the reorganization of the leading modes at zero flux.

The finite-flux enhancement $E$ remains small at low
fields but becomes more pronounced near and beyond the
mode-reorganization region. A narrow enhancement ridge
appears around $B\simeq1.22$--$1.24$. At the
mirror-preserving orientation $\theta_B=0$, this feature
originates from a crossing of the leading modes at zero
flux. A small normal flux mixes these modes, producing two
symmetric maxima of $T_c(\phi)$ at finite flux.
This enhancement is reciprocal and occurs even when
the flux-odd response vanishes. It provides an additional signature of mode reorganization, distinct from the nonreciprocal LP response.

The $\Phi_0$ periodicity follows from gauge invariance,
while the trigonal crystal symmetry constrains the angular
dependence of the flux-odd response. Its magnitude and
field dependence depend on the GL coefficients.
In the present model, the leading cubic contribution arises
from three field-linear couplings connecting winding modes, for example, $n\to n+1\to n+2\to n$.
The first two steps change the winding number by one,
while the last involves a tensorial Lifshitz coupling
that changes it by two in the opposite direction.
Since each coupling is proportional to $B$, their combined
contribution is proportional to $B^3$. Different sequences can contribute with opposite signs, so their contributions may partially or completely cancel.
\begin{figure}[t]
\centering
\includegraphics[width=\columnwidth]{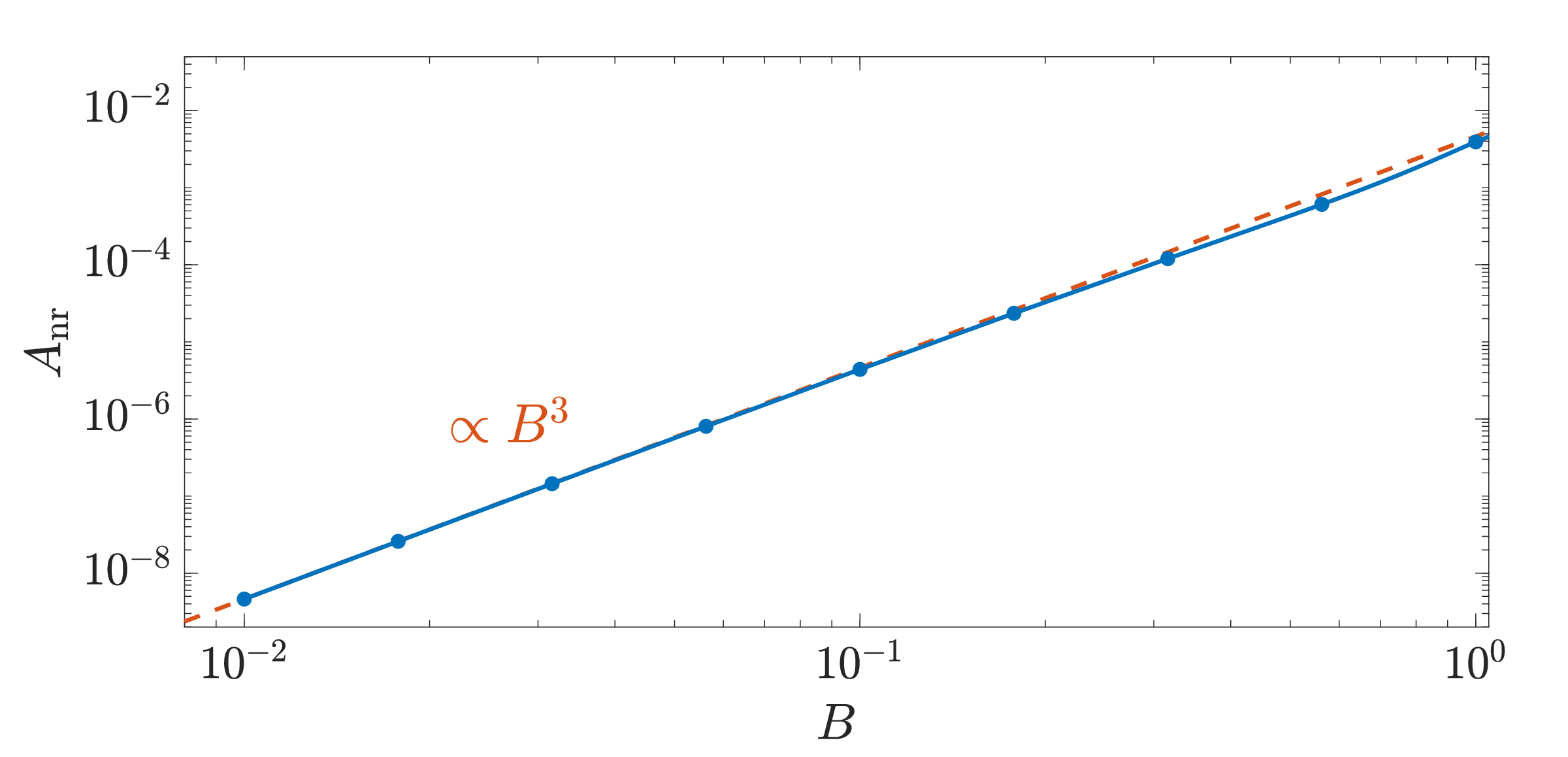}
\caption{Weak-field cubic onset of the nonreciprocal LP response. $A_{\rm nr}(B)$ at $R=1.25$ and $\theta_B=\pi/6$,
shown over two decades in the in-plane field on a log-log
scale. Circles connected by a solid line show the numerical
results and the dashed line is a $B^3$ guide. The response
approaches the cubic dependence at weak fields and deviates
from it toward the upper end of the plotted interval. The suppression associated with mode reorganization occurs
at larger fields, outside this interval (see
Fig.~\ref{fig:Anr_signed_maps}).}
\label{fig:cubic_scaling}
\end{figure}

For the parameters in Table~\ref{tab:gl_params}, setting
$\xi'_{1E}=\xi'_{2E}=0$ reduces $A_{\rm nr}$ below
$10^{-12}$ for all scanned fields and orientations.
This confirms that the tensorial couplings are essential
for the nonreciprocal response within the retained model. The resulting cubic weak-field onset is demonstrated
numerically in Fig.~\ref{fig:cubic_scaling}, while its
perturbative derivation is given in Appendix~\ref{app:analytic_origin}.

The angular dependence is clarified by the signed quantity $O_{\phi_0}$, shown in the center panel of Fig.~\ref{fig:Anr_signed_maps}, which avoids the absolute value in $A_{\rm nr}$ and exposes the sign structure of the nonreciprocal response directly.
The computed signed response obeys the expected magnetic-time-reversal property,
\begin{equation}
O_{\phi_0}(B,\theta_B+\pi)=-O_{\phi_0}(B,\theta_B),
\end{equation}
and is dominated by a single trigonal harmonic,
\begin{equation}
O_{\phi_0}(B,\theta_B)= \mathcal A(B)\cos[3(\theta_B-\theta_m)]+\dots,
\label{eq:numerical_trigonal_fit}
\end{equation}
whose phase is set by the mirror-line angle $\theta_m$: $|O_{\phi_0}|$ is largest for fields parallel to a mirror line and has opposite signs for adjacent mirror lines. Here, $\mathcal A(B)$ is the signed coefficient of the
$m=3$ angular harmonic at the chosen reference flux and radius.
The omitted terms contain the higher allowed angular harmonics.
In fact, the angular dependence is constrained exactly by symmetry (see Appendix~\ref{app:analytic_origin}): $C_{3v}$ invariance, $O_{\phi_0}(B,\theta_B+2\pi/3)=O_{\phi_0}(B,\theta_B)$, together with the field-reversal sign change, $O_{\phi_0}(B,\theta_B+\pi)=-O_{\phi_0}(B,\theta_B)$, restricts the Fourier content to odd multiples of $3$, $m\in\{3,9,15,\dots\}$.  
A discrete Fourier analysis of $O_{1/4}$ shows that the $m=3$ term dominates: over the displayed field range the $m=9$ amplitude stays below $1.4\%$ of the largest $m=3$ amplitude. Only near the polarity reversal at $B\simeq1.26$, where $\mathcal A(B)$ changes sign, does the $m=9$ term become comparable to the $m=3$ term. The same two symmetry relations give $O_{\phi_0}(B,\theta_B+\pi/3)=-O_{\phi_0}(B,\theta_B)$, and the analogous relation holds at every flux. Consequently, the positive amplitude $A_{\rm nr}$ has an exact $60^\circ$ periodicity, whereas the signed response has period $120^\circ$.

\begin{figure}[tb]
\centering
\includegraphics[width=\columnwidth]{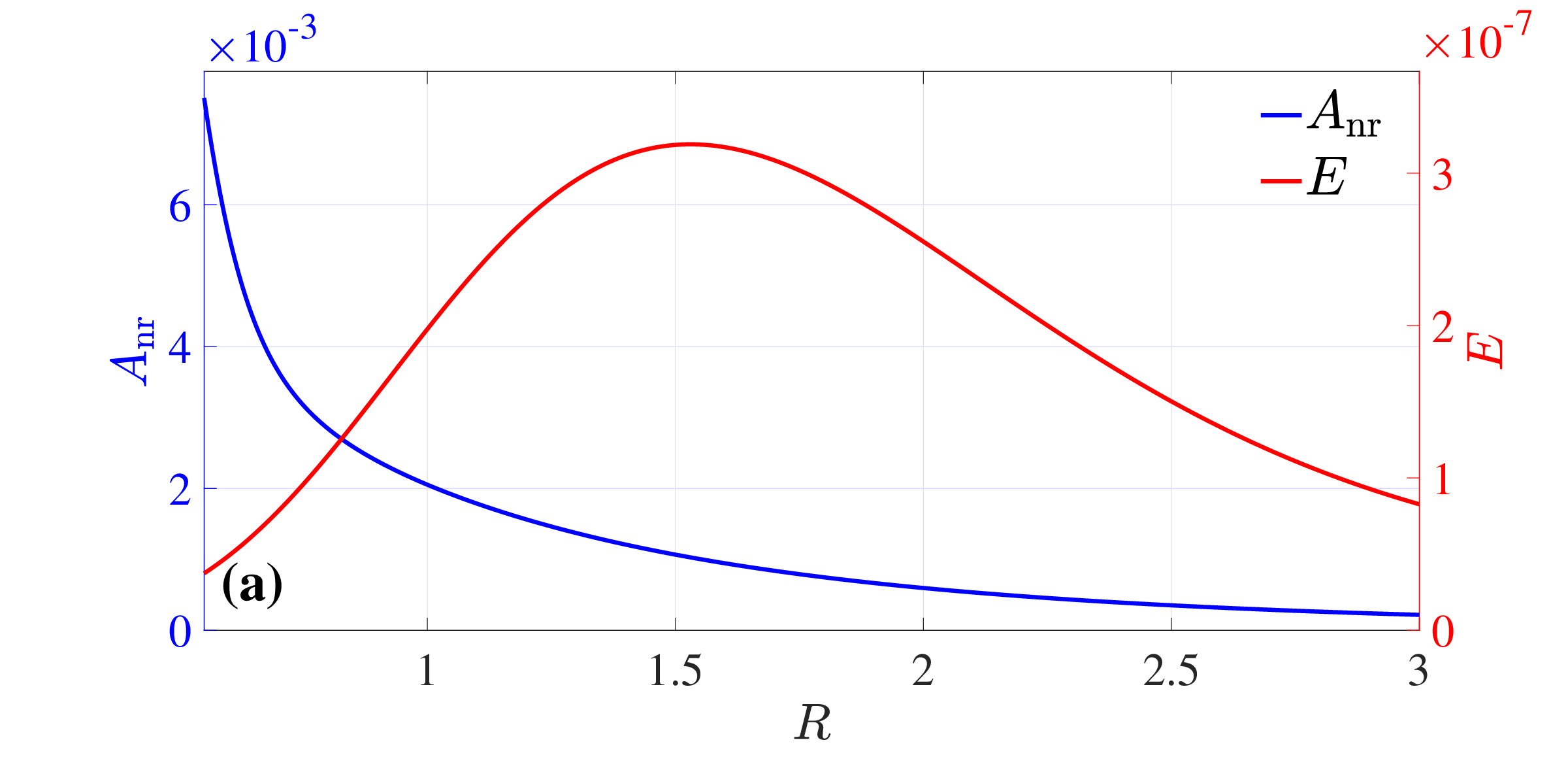}
\includegraphics[width=\columnwidth]{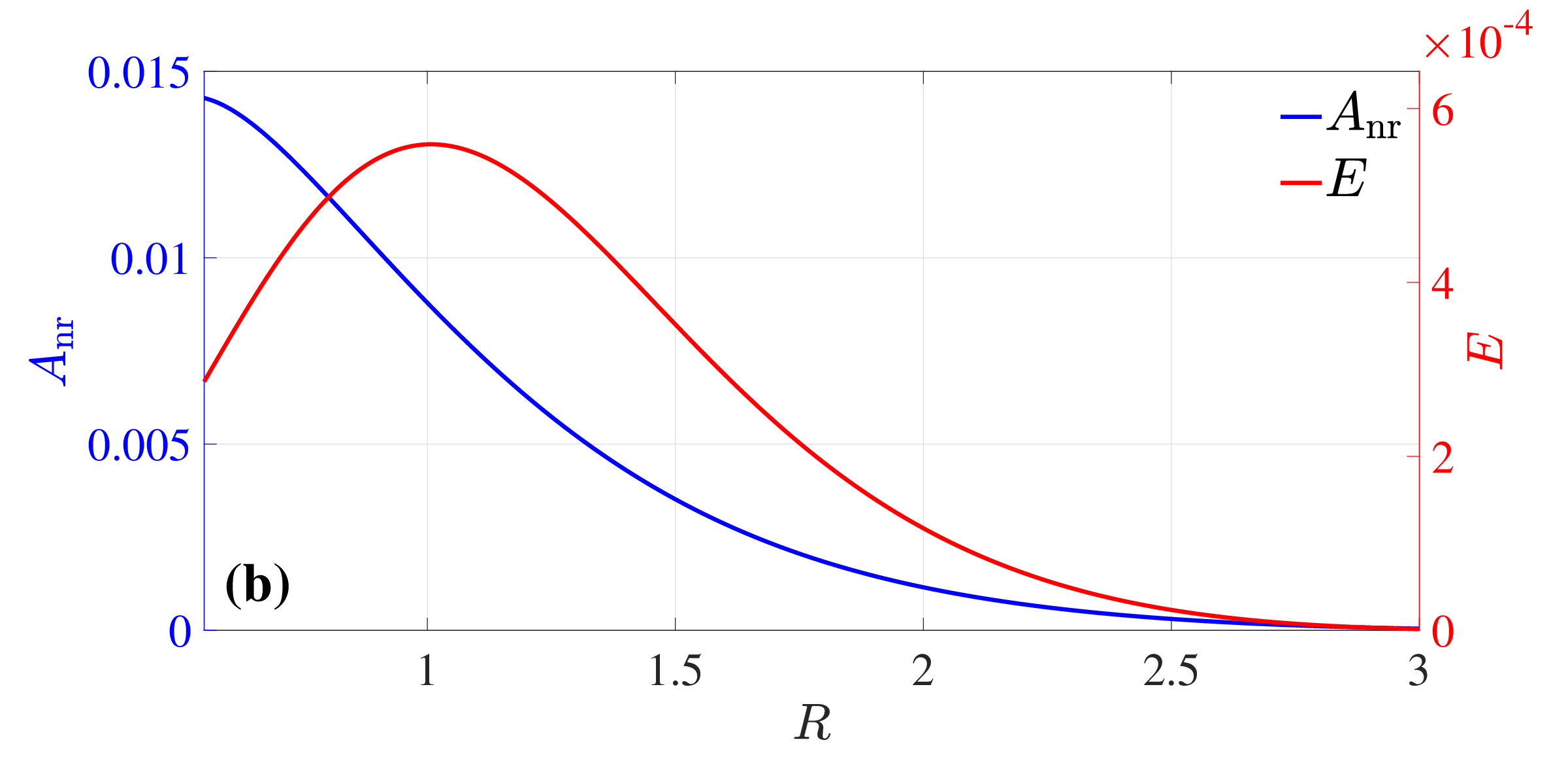}
\caption{Radius dependence of the absolute odd response $A_{\rm nr}$ (blue, left axis) and the finite-flux enhancement $E$ (red, right axis) at $\theta_B=\pi/6$ for (a) $B=0.75$ and (b) $B=2$.}
\label{fig:radius_scan}
\end{figure}

Finally, we study the radius dependence at $\theta_B=\pi/6$ for the two fields of Fig.~\ref{fig:linecut_cartesian}, which lie on either side of the mode reorganization. The odd response and the finite-flux enhancement for $B=0.75$, inside the $A_2$-dominated regime, are shown in Fig.~\ref{fig:radius_scan}(a), while the same quantities for $B=2$, where the leading mode is dominated by the $A_1$ and $E$ components, are shown in Fig.~\ref{fig:radius_scan}(b).
At both fields, the absolute odd response $A_{\rm nr}(R)$ decreases monotonically over the plotted range so that the absolute nonreciprocity is largest at the smallest radii considered. It decays faster at $B=2$ and falls below the $B=0.75$ value for $R\gtrsim2.4$. The two fields differ qualitatively in $E(R)$. At $B=0.75$, see Fig.~\ref{fig:radius_scan}(a), $E(R)$ remains at least three orders of magnitude smaller than $A_{\rm nr}$. At $B=2$, see Fig.~\ref{fig:radius_scan}(b), $E(R)$ has its maximum near $R\simeq1$, where it is about an order of magnitude below $A_{\rm nr}$, and the maximizing flux lies on the opposite side of integer flux from that at $B=0.75$, consistent with the sign reversal between the two fields. Together with the maps in Fig.~\ref{fig:Anr_signed_maps}, this shows that, for the representative parameters, appreciable finite-flux enhancement occurs near and beyond the mode reorganization.

To assess the dependence on the coefficients, we have repeated the calculation for modified parameter sets that change the coefficients of Table~\ref{tab:gl_params} by $\pm20\%$, added the channel-dependent pair breaking
\begin{equation}
\delta\mathcal F^{(2)}=B^2\bigl[\beta_1|\psi_1|^2+\beta_2|\psi_2|^2+\beta_E\bigl(|\psi_x|^2+|\psi_y|^2\bigr)\bigr],
\label{eq:pair_breaking}
\end{equation}
or combined both modifications, (for details see Appendix~\ref{app:cartesian_projection}). The real coefficients $\beta_a\geq0$ describe channel-resolved
pair breaking in the dimensionless normalization used here.
For all modifications, the leading instability at weak field remains dominated by the $A_2$ component and the odd response persists, while its amplitude varies by about an order of magnitude. The field of the mode reorganization also depends on the coefficients. For some sets the dominance of $A_2$ is already lost at $B=0.75$.

For the parameters in Table~\ref{tab:gl_params}, the weak-field response has the form $O_{1/4}=c_3B^3+\mathcal O(B^5)$.
The cubic coefficient contains products of three field-linear couplings: one non-derivative, one tensorial, and one helical Lifshitz coupling. For example, a contribution can be proportional to $\zeta_{2E}\xi'_{2E}\xi_2$. Different contributions can have opposite signs and may cancel, so neither the magnitude nor the sign of $c_3$ is universal.

Such cancellations can suppress the cubic term without eliminating the response at higher orders. In contrast, the entire odd response vanishes at mirror-preserving field orientations or, for the retained functional with $\eta_{1E}=\eta_{2E}=0$, when one of these three groups of couplings is absent. Further details are given in
Appendix~\ref{app:analytic_origin}.

The parameters in Table~\ref{tab:gl_params} do not include
pair breaking by the in-plane field. Consequently, the
field-linear couplings can increase $T_c$ by mixing
pairing channels. Our results near
and beyond the mode reorganization therefore assume
that superconductivity survives up to these fields.

For the equal temperature slopes $a_a$ used here,
the channel-independent pair breaking, described by
Eq.~\eqref{eq:pair_breaking} with $\beta_1=\beta_2=\beta_E$, shifts $T_c$ downward without changing its flux dependence at fixed fields (see Appendix~\ref{app:cartesian_projection}).

If the field-linear couplings originate from the
Zeeman effect, their natural field scale corresponds
to Zeeman energies of the order of $k_BT_c$. For a reference
field $B_{\rm ref}$ of this order, the mode reorganization
may occur near the Pauli limit, while the largest fields considered here may exceed it. Strong spin-momentum locking can weaken the paramagnetic pair breaking and favor helical superconductivity under an in-plane field~\cite{Dimitrova2007,Smidman2017}.

Importantly, the cubic weak-field response does not
require superconductivity to persist up to the larger fields where the modes reorganize.

% ============================================================
\section{Discussion}
\label{sec:discussion}
% ============================================================

Our numerical results are based on the representative parameter set in Table~\ref{tab:gl_params} and the variations discussed above. We have not attempted an exhaustive exploration of the GL parameter space. Our calculation determines the nucleation temperature $T_c(\phi,\mathbf B_\parallel)$ and the corresponding
order-parameter mode. Determining the superconducting phases below this temperature would require inclusion of the quartic GL terms.

Field-induced competition between pairing channels has also been investigated in the related layered compound $\beta$-PdBi$_2$, where tunnelling measurements provide evidence for a change from fully gapped to nodal superconductivity under an in-plane field. This behaviour was interpreted in terms of competing $s$- and $p$-wave states~\cite{Powell2025}. The proposed transition occurs
within the superconducting state and is distinct from the field evolution of the nucleating mode studied here.

The LP response constrains the pairing state without identifying it. In the homogeneous ideal ring considered here, a single-component order parameter gives an exactly reciprocal LP curve, regardless of its representation and singlet-triplet content (see Appendix~\ref{app:analytic_origin}). Within this setting, the flux-odd response requires coupled pairing channels. In the present functional, these are a scalar channel and the $E$ doublet channel, coupled by the tensorial invariants. Additional boundary or spatial structure of a real ring can relax this requirement. The LP signatures do not uniquely identify an $A_2$ parent state since an $A_1$ or an $E$ parent state with the same couplings gives a response of comparable or larger (see Appendix~\ref{app:analytic_origin}). Pair breaking of the form of Eq.~\eqref{eq:pair_breaking}, through which Zeeman effects on singlet and triplet components differ, leaves the coefficient of the cubic $B^3$ dependence of the odd response $O_{1/4}$ unchanged. The symmetry of the response is fixed by the crystal structure, whereas its sign and magnitude depend on the coefficients and on edge contributions (see Appendix~\ref{app:cartesian_projection}).

While symmetry constrains the allowed nonreciprocal
LP response, its magnitude, sign, and characteristic field scales depend on the GL coefficients and the ring radius. Quantitative comparison with future experiments on PtBi$_2$ rings will  require device-specific parameters.
We predict a signed, field-orientation-controlled, flux-odd transition-temperature component generated by the retained multicomponent couplings. Finite-flux enhancement and shifted extrema are additional, more parameter-dependent features. Taken together, these features constitute a testable scenario of the proposed multicomponent magnetoelectric mechanism. They are compatible with an $A_2$ ($i$-wave) parent state in PtBi$_2$ but do not by themselves uniquely identify the order-parameter symmetry. The calculated LP response does not determine the relative spin-singlet and spin-triplet contributions within the pairing channels. Since the GL functional is formulated in terms of these representation amplitudes the singlet and triplet contributions enter only through the values of its coefficients and the present transition-temperature calculations do not determine their relative weight. 

LP interferometry would complement existing probes of surface
superconductivity in PtBi$_2$. Angle-resolved photoemission
constrains the superconducting Fermi arcs and their momentum-resolved gap structure~\cite{Kuibarov2024,Changdar2025}, while tunnelling spectroscopy, vortex imaging, and Josephson measurements provide local spectral information and evidence of phase coherence~\cite{Moreno2026}. The proposed LP experiment addresses a different aspect: the response of the superconducting instability to phase winding and independently controlled magnetic fields. The combined flux, field-angle, and radius dependencies could constrain field-induced coupling between pairing channels, including channels that remain subdominant at zero field. For the parameters studied here, the nonreciprocal response tends to zero as $B^3$ when the in-plane
field vanishes. The larger amplitudes found at higher
fields are associated with a leading instability
dominated by the $A_1$ and $E$ components rather than the zero-field $A_2$ parent. This information is complementary to
%, rather than a substitute for,
identifying the momentum-space nodes or gap-sign structure.
Provided the superconducting state of the patterned device is
independently characterized, comparison with spectroscopic
constraints could test whether a common multicomponent description accounts for both the gap structure and the mesoscopic magnetic response.
We stress that the calculated enhancement of superconductivity at weak and intermediate in-plane fields
is accompanied by an increased admixture of the $A_1$
and $E$ components to the initial $A_2$ parent state.
To the best of our knowledge, such mesoscopic tuning
of an unconventional multicomponent superconductor
by external magnetic fields, accompanied by the nonreciprocal LP response described here, has not been reported previously.

Temperature-dependent scanning SQUID microscopy could
provide a contactless route to investigate the
flux-dependent superconducting onset at fixed in-plane
field, complementing transport
measurements~\cite{Koshnick2007}.
Scanning SQUID measurements have already detected a
diamagnetic response in PtBi$_2$~\cite{Schimmel2024}.
Relating the measured magnetic response to the calculated
mean-field $T_c$ would require accounting for
superconducting fluctuations, which can be significant
in narrow rings near the transition~\cite{Bert2011}.

For the representative parameters studied here, the in-plane magnetic field can raise the calculated nucleation temperature through mixing of pairing channels ~\cite{remark3}. This mechanism, explained above in terms of field-tuned multicomponent superconductivity in the mesoscopic rings, differs from the exchange-field compensation underlying the Jaccarino--Peter effect~\cite{JaccarinoPeter1962}. Whether a net increase occurs in an experimental device, particularly at intermediate and high in-plane magnetic fields, depends on the competition between the enhancement considered here and orbital and paramagnetic suppression of superconductivity by the same field.
Although Fig.~\ref{fig:mode_content} shows that the
leading instability becomes dominated by the $A_1$
and $E$ channels, this does not establish spin-triplet
dominance or survival beyond the conventional Pauli limit.

Related GL studies of Rashba superconductors illustrate how spin--orbit coupling can modify the magnetic-field response~\cite{Kuzmartsev2026} with respect to the upper critical field pecularities in case of bulk superconductivity. Neither surface superconductivity nor the LP effect have been addressed in the literature to the best of our knowledge, so far. Anyhow, despite the still necessary increase of the already large number of parameters, a useful extension of the present approach could explicitly resolve the spin-singlet and spin-triplet contributions within the pairing channels, respecting the $C_{3v}$ symmetry of the PtBi$_2$ surface. Such a treatment could clarify their influence on the magnetic response of the proposed $A_2$ ($i$-wave) state.

 A self-consistent treatment of surface and bulk superconductivity in time-reversal-symmetric Weyl semimetals shows how the penetration of Fermi-arc states into the bulk influences the surface gap and its temperature dependence~\cite{Trama2025}. Microscopic proposals for surface superconductivity include phonon-mediated pairing on Fermi arcs~\cite{Maeland2025}. For PtBi$_2$, an $i$-wave instability has also been obtained within an electronic Kohn--Luttinger approach~\cite{Dsouza2026}, while anisotropic electron--phonon coupling combined with statically screened Coulomb repulsion provides another proposed
route to nodal pairing~\cite{Maeland2026}.
In the latter study, five distinct gap functions emerge in different regions of the interaction-parameter space, including the nodal $i$-wave profile considered here.
Our phenomenological GL analysis does not select between these microscopic mechanisms. It studies the LP response of coupled symmetry-allowed pairing channels.

%============================================================
\section{Conclusions}
\label{sec:conclusions}
% ============================================================

We have developed a multicomponent GL theory for the LP effect for a narrow ring with $A_1\oplus A_2\oplus E$ surface order-parameter sectors, motivated by the symmetry structure proposed for PtBi$_2$. 
The resulting LP response remains strictly $\Phi_0$-periodic but at fixed in-plane field it can be nonreciprocal in the flux through the ring. At weak in-plane fields, the leading symmetry-allowed flux-odd response is proportional to the cube of the field strength. It follows the threefold crystal
symmetry and changes sign when the field direction
is rotated by $\pi/3$. Its periodic flux dependence can contain several sine harmonics. The GL functional produces a nonreciprocal response through the combined action of couplings between modes whose winding numbers differ by one or two, with the latter arising from tensorial Lifshitz terms. Because processes of opposite sign compete, the sign and size of the cubic contribution depend on the GL coefficients.
For the representative coefficients, the in-plane field also changes the character of the leading instability, from predominantly $A_2$ at weak field to predominantly $A_1$ and $E$, where the larger odd amplitudes obtained here belong to the field-reorganized mode rather than to the low-field $A_2$ state.

Finite-flux enhancement is possible but is more parameter dependent. For the representative parameters, appreciable enhancement occurs near and beyond the mode reorganization. Furthermore, we find that the odd-in-flux response decreases with increasing radius for both the field values we considered. Hence, the calculation does not suggest an optimal intermediate radius for the nonreciprocal response.

An experimental test should combine flux reversal, in-plane-field reversal, and crystallographically referenced angular scans with independent calibration of field misalignment and flux offsets. Reversal tests alone would not uniquely distinguish the proposed mechanism from geometrical asymmetry~\cite{Vekris2021}.
 
Because the coefficients used here are representative rather than fitted, a quantitative comparison with PtBi$_2$ devices requires estimates of the surface GL coefficients, which set the magnitude of the odd response and the field scale. It also requires relating the calculated $T_c(\phi,\mathbf B_\parallel)$, the mean-field temperature at which superconductivity nucleates, to the measured resistance. Converting resistance oscillations into $T_c$ oscillations through the temperature slope of the resistance assumes that the resistive transition shifts rigidly with $T_c$. However, PtBi$_2$ flakes show a BKT transition~\cite{BKT_PtBi2} and in such a two-dimensional superconductor the resistance between the BKT and mean-field transitions arises from thermally excited free vortices. Near the BKT transition, its flux dependence can follow the free-vortex density rather than $T_c(\phi,\mathbf B_\parallel)$ and, depending on the ring width, can even be shifted by half a flux quantum~\cite{XieNagaosa2026}.

Nevertheless, our results provide a concrete  and timely motivation for experimental LP studies of mesoscopic PtBi$_2$ rings with independently controlled out-of-plane flux and in-plane magnetic field. The predicted flux asymmetry, cubic weak-field onset, and crystallographically locked angular dependence offer testable consequences of multicomponent magnetoelectric coupling. In the representative parameter regime, the field-induced sign reversal and the distinct radius dependences of nonreciprocity and finite-flux enhancement provide further tests of the evolution of the superconducting instability. Such measurements could constrain the coupling between competing pairing channels and complement spectroscopic investigations of the proposed nodal surface state.
The thin-ring LP approach developed here for the $i$-wave scenario in $\gamma$-PtBi$_2$ may also be used to investigate other unconventional surface pairing states within multicomponent GL models. A systematic comparative study of the corresponding GL functionals is left for future work.

\begin{acknowledgments}
We thank Marieke Altena, Flavio S. Nogueira, Ulrich~R\"o{\ss}ler and Sergey Borisenko for fruitful discussions. We acknowledge financial support by the Deutsche Forschungsgemeinschaft (DFG) through the Collaborative Research Center SFB 1143, projects A04 and A05, Project ID 247310070, and the W\"urzburg-Dresden Cluster of Excellence on Complexity, Topology and Dynamics in Quantum Matter---ctd.qmat, EXC 2147, Project ID 390858490.
\end{acknowledgments}

% ============================================================
\appendix

% ============================================================
\section{Additional remarks on the quadratic functional}
\label{app:quadratic_invariants}
% ============================================================

Here, we discuss the choice of quadratic terms in Sec.~\ref{sec:GL}, their relation to alternative invariant
forms, and the boundary contributions relevant to the effective ring model.

The $C_{3v}$ symmetry also permits the symmetric traceless
$E$-sector gradient term $\gamma''_E\bigl(|C_1|^2+|C_2|^2\bigr)$.
Under the tangential derivative replacement used in
Sec.~\ref{sec:ring}, the identity
\begin{equation}
|C_1|^2+|C_2|^2=|\Sigma|^2+|\Delta|^2
\end{equation}
holds pointwise. This term can be absorbed into the effective ring stiffnesses through
\begin{equation}
\gamma_E\longrightarrow\gamma_E+\gamma''_E,
\qquad
\gamma'_E\longrightarrow\gamma'_E+\gamma''_E.
\end{equation}
We set $\gamma''_E=0$ after incorporating its contribution into $\gamma_E$ and $\gamma'_E$. This redundancy is specific to the tangential model and does not establish equivalence between general finite-width annuli with different boundary conditions.

The zero-field mixed Lifshitz terms in Eqs.~\eqref{eq:app_F1EL} and \eqref{eq:app_F2EL} are understood under the spatial integral. They differ from the real parts of the alternative
forms in Ref.~\cite{Timm2025}, $\eta_{1E}\bigl(\psi_1^*\Sigma
-\boldsymbol\psi_E^*\cdot\mathbf D\psi_1\bigr)$ and its $A_2$ counterpart, only by total derivatives in two dimensions. Integration by parts leaves the functional unchanged when the resulting boundary contributions are included. For a finite-width
annulus, azimuthal periodicity removes boundary contributions along the circumference, but not those at the inner and outer edges. Applying the tangential derivative replacement to different Cartesian forms without retaining the corresponding boundary terms can produce different curvature couplings
of order $1/R$ in the effective ring model.

For example, Eqs.~\eqref{eq:app_F1EL} and \eqref{eq:app_F2EL} give the winding-preserving matrix elements
\begin{equation}
\mathbb K_{1n,rn}=\frac{\eta_{1E}}{R},
\qquad
\mathbb K_{2n,tn}=\frac{\eta_{2E}}{R},
\end{equation}
whereas applying the same tangential prescription to the real parts of the alternative forms gives half these values. We use the expressions specified in Sec.~\ref{sec:GL} as written. Since
$\eta_{1E}=\eta_{2E}=0$ in all calculations reported here, this particular difference does not affect the numerical results.

A similar dependence on the boundary prescription arises for the intercomponent contributions to the field-linear gradient terms in Eqs.~\eqref{eq:app_F12L}--\eqref{eq:app_F2ELprime}.
These couplings are nonzero in the numerical calculations. Changing their Cartesian forms without including the accompanying boundary contributions can modify the magnitude and polarity of the flux-odd response, as discussed in
Appendix~\ref{app:cartesian_projection}. Such changes correspond to different effective ring functionals, rather than equivalent rewritings of the same complete functional.

Additional symmetry-allowed intersector cross-gradient
terms, including those in Ref.~\cite{Timm2025}, are
set to zero here. In the ring representation, such terms can introduce zero-field couplings with $\Delta n=\pm3$. These allow additional closed coupling sequences involving three field-induced couplings, each proportional to $B$, beyond those
discussed for the present functional. Consequently, the necessity of the tensorial $\Delta n=\pm2$ couplings for the nonreciprocal response is a property of the specified model,
not of every $C_{3v}$-symmetric extension. The symmetry constraints on the allowed angular dependence remain valid for such symmetry-preserving
extensions.

% ============================================================
\section{Ring reduction and numerical implementation}
\label{app:cartesian_projection}
% ============================================================

Here, we describe how the Cartesian GL functional is
reduced to the effective one-dimensional ring model
and represented in the Fourier basis. We assume that
the order parameters depend only on the angular coordinate and express the Cartesian derivatives in terms of tangential derivatives. The resulting differential operators are  projected onto the winding modes $e^{in\theta}$ to construct the matrix used in the numerical calculations. The starting point is the quadratic functional specified in Sec.~\ref{sec:GL}.  For each angle $\theta$, we express the components $(\psi_x,\psi_y)$ through $(\psi_r,\psi_t)$ using Eq.~\eqref{eq:cart_local_basis}. The angular derivatives of the basis vectors are retained throughout.  The GL matrix is obtained from the Fourier matrix element
\begin{equation}
\mathbb K_{\mu n,\nu m}
=
\frac{1}{2\pi}
\int_0^{2\pi}d\theta\,
 e^{-in\theta}\,
\widehat{\mathcal K}_{\mu\nu}(\theta)\,
 e^{im\theta},
\label{eq:app_projection_again}
\end{equation}
where $\mu,\nu\in\{1,2,r,t\}$.  In practice this integral can be evaluated either analytically or by high-resolution Fourier quadrature.  The latter is particularly useful because it avoids premature simplification of the tensorial Lifshitz terms.

Carrying out the angular reduction explicitly, the local basis vectors $\mathbf n=(\cos\theta,\sin\theta)$ and $\mathbf t=(-\sin\theta,\cos\theta)$ satisfy $\partial_\theta\mathbf n=\mathbf t$ and $\partial_\theta\mathbf t=-\mathbf n$, so that the divergence- and curl-like $E$-sector gradients reduce to
\begin{align}
\Sigma &\equiv D_x\psi_x+D_y\psi_y\longrightarrow D_s\psi_t+\frac{1}{R}\psi_r, \\
\Delta &\equiv D_x\psi_y-D_y\psi_x\longrightarrow -D_s\psi_r+\frac{1}{R}\psi_t,
\end{align}
while the symmetric traceless combinations $C_1=D_x\psi_y+D_y\psi_x$ and $C_2=D_x\psi_x-D_y\psi_y$ follow from
\begin{equation}
D_+\psi_+=C_2+iC_1\longrightarrow e^{2i\theta}\left(iD_s-\frac{1}{R}\right)(\psi_r+i\psi_t),
\end{equation}
with $\psi_+=\psi_x+i\psi_y=e^{i\theta}(\psi_r+i\psi_t)$ and $D_+=D_x+iD_y$.  The $e^{2i\theta}$ factor is the origin of the $\Delta n=\pm2$ blocks, and the curvature-shifted operator reduces on winding sector $n$ to the real factor $-k_{n+1}$ of Eq.~\eqref{eq:kappa_real_main}.

The field-dependent invariants have the generic form
\begin{equation}
iB_j\left[X_j-X_j^*\right].
\label{eq:app_iBXX}
\end{equation}
Both terms in Eq.~\eqref{eq:app_iBXX} are retained. If $M_j$ denotes the Fourier projection of $X_j$, the projection of $X_j^*$ is $M_j^\dagger$, and for real $B_j$ the contribution to the GL matrix is
\begin{equation}
\mathbb K_j=iB_j\bigl(M_j-M_j^\dagger\bigr)=iB_jM_j+\bigl(iB_jM_j\bigr)^\dagger.
\label{eq:app_Kj}
\end{equation}
Projecting the complete term $iB_jX_j$ and adding its Hermitian conjugate, including the factor $iB_j$, is equivalent to projecting the full expression. For invariants that couple different components forms related by integration by parts differ on the narrow ring by contributions from its inner and outer edges, which are of order $1/R$, like the curvature terms that produce the odd response. We use the invariant forms specified in Sec.~\ref{sec:GL}, with the boundary prescription
discussed in Appendix~\ref{app:quadratic_invariants}. Other choices leave the symmetry of the response unchanged but can change its magnitude and reverse its weak-field polarity.

The resulting matrix has only $\Delta n=0,\pm1,\pm2$ blocks:
\begin{equation}
\mathbb K_{nm}=0
\qquad\text{for}\qquad |n-m|>2.
\end{equation}
We denote
\begin{equation}
A_n=\mathbb K_{n,n},
\quad
H^{(1)}_n=\mathbb K_{n+1,n},
\quad
H^{(2)}_n=\mathbb K_{n+2,n}.
\label{eq:app_AH1H2_def}
\end{equation}
The full truncated matrix is reconstructed from these blocks as
\begin{equation}
\begin{split}
\mathbb K_N={}& \sum_n A_n +
\sum_n\left(H^{(1)}_n+H^{(1)\dagger}_n\right)\\
&+ \sum_n\left(H^{(2)}_n+H^{(2)\dagger}_n\right),
\end{split}
\label{eq:app_block_reconstruct}
\end{equation}
where the notation indicates the corresponding block placement in winding space.

For the reference projection used to validate the implementation, the GL matrix is Hermitian by construction, so the nontrivial check is magnetic time reversal.  Building the matrix independently at $(\phi,\mathbf B_\parallel)$ and at $(-\phi,-\mathbf B_\parallel)$ and comparing them, we find that Eq.~\eqref{eq:kernel_TR} is satisfied with internal action $U_{\mathcal T}=I_4$ to machine precision.  This is the expected result because in the original GL basis the order parameters transform to their complex conjugates under time reversal.  In the real Cartesian and radial--tangential component bases used here, no additional internal sign matrix is required. A different internal basis can have a nontrivial representation of time reversal and the representation must be transformed consistently with the GL matrix.

Unless stated otherwise, the calculations in Sec.~\ref{sec:results} use the winding truncation $N=15$, which is sufficient for the displayed parameter sets, and evaluate the
angular Fourier integrals with $M_\theta=1024$ uniformly spaced quadrature points. For the cubic-scaling results in Fig.~\ref{fig:cubic_scaling}, we use $N=12$, $M_\theta=512$, and 401 flux points over $\phi\in[-1/2,1/2]$.
The field-angle maps use $\phi\in[-1/2,1/2]$ with 801 points, $B\in[0,2.5]$ with 61 points, and $\theta_B\in[-\pi,\pi]$ with 121 points. The radius scans at $B=0.75$ and $B=2$ use $R\in[0.55,3]$ with 301 points each and 1001 flux points. In the radius scans, $E$ is obtained by refining the maximum of $T_c(\phi)$ with a bounded one-dimensional search between the grid points adjacent to the grid maximum. This matters at weak field: at $B=0.75$ and $R\lesssim1$ the maximizing flux lies closer to integer flux than half the grid spacing $10^{-3}$, so that the grid value of $E$ vanishes there although $E$ itself is small but finite. In these calculations the relative residual of Eq.~\eqref{eq:kernel_TR} stays below $10^{-12}$, and the $\Phi_0$-periodicity error is negligible on the scale of the plotted data.

Winding-cutoff and angular-quadrature convergence were checked separately from the resolution of the maximum over flux.

The sensitivity test of Sec.~\ref{sec:results} uses 56 parameter sets. Twelve sets change one group of coefficients (non-derivative, helical, tensorial, or stiffness) or one bare channel separation by $\pm20\%$. Twelve sets add the pair breaking of Eq.~\eqref{eq:pair_breaking} with $(\beta_1,\beta_2,\beta_E)$ running over the permutations of $(0.05,0.10,0.15)$ and $(0.10,0.20,0.30)$. The remaining 32 sets combine independent $\pm20\%$ changes of the nonzero field-linear coefficients, stiffnesses, and channel separations with $0\leq\beta_a\leq0.2$. The bare $A_2$ transition temperature and the mass coefficients are kept at unity, and the coefficients that vanish in Table~\ref{tab:gl_params} remain zero. All sets are evaluated at $R=1.25$ and $\theta_B=\pi/6$. At $B=0.5$, all of them keep $w_2>0.6$ over the whole flux period, and $A_{\rm nr}$ ranges from $0.07\%$ to $0.76\%$ of the LP amplitude, compared with $0.33\%$ for Table~\ref{tab:gl_params}. At $B=0.75$, 14 sets lose the $A_2$ majority somewhere in the flux period. At $B=2$, the twelve pair-breaking sets keep $T_c\geq0.6$ over the whole flux period, whereas channel-independent pair breaking suppresses $T_c$ to zero at this field for $\beta\gtrsim0.34$.

% ============================================================
\section{Symmetry, mechanism, and scale of the nonreciprocal Little--Parks response}
\label{app:analytic_origin}
% ============================================================

The weak-field power counting used below assumes an analytic expansion of an isolated instability eigenvalue at fixed flux. At winding-sector or internal-mode degeneracies, degenerate perturbation theory or direct diagonalization is required. In particular, the fixed-flux expansion need not be uniform near a branch degeneracy.

The transition temperature $T_c(\phi,\mathbf B_\parallel)$ is a real scalar and must be invariant under the magnetic symmetry of the surface: the unitary operations of $C_{3v}$, supplemented by time reversal.  The normal flux $\phi$ transforms like the perpendicular field component $B_z$, independently of the field profile: it is invariant under the threefold rotation and odd under a vertical mirror, and therefore transforms as $A_2$.  The in-plane field transforms as an $E$ doublet.  The nonreciprocal part of the LP response is odd under $\phi\to-\phi$, and hence its leading local dependence on flux transforms as $A_2$.  To obtain an invariant scalar, this $A_2$ factor must multiply an $A_2$ polynomial of the in-plane field.

Writing the in-plane field as $B_\pm=B_x\pm iB_y=Be^{\pm i\theta_B}$, its transformation under the generators of $C_{3v}$ is fixed.  The threefold rotation acts as $\theta_B\to\theta_B+2\pi/3$,
\begin{equation}
C_3:\quad B_\pm\to e^{\pm 2\pi i/3}\,B_\pm,
\label{eq:app_C3_action}
\end{equation}
while, in a frame whose $x$ axis lies along a mirror line ($\theta_m=0$), the corresponding vertical mirror reverses the chirality of the in-plane field,
\begin{equation}
\sigma_v:\quad B_\pm\to -\,B_\mp=-\,B_\pm^{*}
\label{eq:app_mirror_action}
\end{equation}
(an arbitrary mirror orientation is restored by the shift $\theta_B\to\theta_B-\theta_m$).  Since the flux-odd factor transforms as $A_2$, we need a polynomial in the field that is invariant under Eq.~\eqref{eq:app_C3_action} and changes sign under Eq.~\eqref{eq:app_mirror_action}; enumerating the symmetric powers of $B_\pm$ locates the lowest one.

At first order, $\{B_+,B_-\}$ transforms as the $E$ doublet.
We denote by $\operatorname{Sym}^{p}E$ the $p$th symmetric
tensor power of this representation. Here, this means the
representation carried by all homogeneous polynomials of
degree $p$ in the field components $B_x,B_y$, or equivalently
$B_+,B_-$. Only symmetric products contribute because the
components of the applied magnetic field commute: changing
the order of the factors does not produce a distinct
polynomial. At second order, these products decompose as
\begin{equation}
\mathrm{Sym}^2E:\quad
\underbrace{B_+B_-}_{A_1}\ \oplus\ \underbrace{\{B_+^2,B_-^2\}}_{E},
\end{equation}
where $B_+B_-=B^2$ is the rotation- and mirror-invariant modulus.  There is no $A_2$: the antisymmetric pairing $B_xB_y-B_yB_x$ that would carry it vanishes identically for a single field.  At third order,
\begin{equation}
\mathrm{Sym}^3E:\quad
\underbrace{\{B_+^3,B_-^3\}}_{A_1\oplus A_2}\ \oplus\ \underbrace{B^2\{B_+,B_-\}}_{E},
\end{equation}
and the cubic pair $\{B_+^3,B_-^3\}$ finally supplies a one-dimensional $A_2$.  Under $C_3$ a cubic monomial accumulates a full rotation and is invariant, $B_+^3\to e^{2\pi i}B_+^3=B_+^3$; under $\sigma_v$ it maps to $B_+^3\to-(B_+^3)^{*}=-B_-^3$.  Writing $B_+^3=B^3(\cos3\theta_B+i\sin3\theta_B)$, the real and imaginary parts therefore transform as
%\begin{equation}
\begin{align}
\operatorname{Re}B_+^3&=B^3\cos3\theta_B\quad (A_2),\\
\operatorname{Im}B_+^3&=B^3\sin3\theta_B\quad (A_1),
\end{align}
%\end{equation}
since the mirror sends $\cos3\theta_B\to-\cos3\theta_B$ while leaving $\sin3\theta_B$ invariant.  The lowest-degree field polynomial transforming as $A_2$ is thus cubic, and for a general mirror orientation it reads
\begin{equation}
B^3\cos3(\theta_B-\theta_m)=\operatorname{Re}\!\left[e^{-3i\theta_m}B_+^3\right].
\label{eq:app_A2_cubic}
\end{equation}
Thus, the leading smooth harmonic of the odd LP response has the form
\begin{equation}
T_c^{\rm odd}(\phi,\mathbf B_\parallel)
\simeq
c_3 B^3 \cos 3(\theta_B-\theta_m)\sin(2\pi\phi),
\label{eq:app_symmetry_form}
\end{equation}
where
\begin{equation}
T_c^{\mathrm{odd}}(\varphi,\mathbf B_\parallel)
=
\frac{T_c(\varphi,\mathbf B_\parallel)
      -T_c(-\varphi,\mathbf B_\parallel)}{2}.
\end{equation}
More generally, each flux harmonic $\sin(2\pi\ell\phi)$ is separately odd in $\phi$ and also transforms as $A_2$, so flux periodicity permits the same cubic angular factor to multiply any of them,
\begin{equation}
\begin{split}
T_c^{\rm odd}(\phi,\mathbf B_\parallel)
={}&
B^3\cos3(\theta_B-\theta_m)\\
&\times\sum_{\ell\ge1} d_\ell \sin(2\pi\ell\phi)
+\cdots,
\end{split}
\end{equation}
where the real coefficients $d_\ell$ label flux harmonics and depend on the radius and the GL coefficients.
Eq. \eqref{eq:app_symmetry_form} should be understood as the leading smooth harmonic rather than as an exact representation of the branch-selected LP curve.  Magnetic time reversal is automatic: $\phi\to-\phi$ and $\theta_B\to\theta_B+\pi$, so both factors in Eq.~\eqref{eq:app_symmetry_form} change sign and their product is invariant.  This gives
\begin{equation}
T_c(\phi,\mathbf B_\parallel)=T_c(-\phi,-\mathbf B_\parallel),
\end{equation}
while still allowing
\begin{equation}
T_c(\phi,\mathbf B_\parallel)\neq T_c(-\phi,\mathbf B_\parallel)
\end{equation}
at fixed in-plane field.

As a result the angular harmonic $m=3$ is symmetry enforced at leading order.  Higher powers of $B$ can renormalize the same $m=3$ harmonic through additional even powers of $B$.  Genuine higher angular harmonics compatible with the same magnetic symmetry occur only at odd multiples of three, $m=9,15,\ldots$.  This explains why the signed numerical map is dominated by a trigonal sign-changing pattern, whereas the absolute-value map $A_{\rm nr}$ displays six lobes.

Symmetry fixes the form of Eq.~\eqref{eq:app_symmetry_form}, but not the coefficient $c_1$.  The mechanism producing a nonzero coefficient within the retained GL model is the winding-space structure of its ring-projected GL matrix.  In the minimal cubic process, the $A_2$ field invariant is produced by three field couplings of the same angular chirality, giving $B_\pm^3$.  In the ring projection each winding step carries a definite field chirality: the field-linear and helical $\Delta n=\pm1$ processes carry $B_\mp$, while the tensorial field-dependent Lifshitz invariants $F^{(L)\prime}_{1E}$ and $F^{(L)\prime}_{2E}$ generate $\Delta n=\pm2$ steps carrying $B_\pm$. Consider three field-induced couplings that are all
proportional to $B_+$ or all to $B_-$. To return to the initial winding sector, their winding changes must sum to zero, $\sum\Delta n=0$. In the present model, the possible combinations are
\begin{equation}
\{+1,+1,-2\},
\qquad
\{-1,-1,+2\},
\label{eq:closed_winding_sequences}
\end{equation}
up to permutations of the three steps.
Each of these triangular processes contains a tensorial $\Delta n=\pm2$ step. Let $\mathbb K_{N,{\rm off}}$ denote the truncated GL matrix with $\xi'_{1E}=\xi'_{2E}=0$. Within the given functional, removing both tensorial coefficients eliminates the flux-odd response exactly, not only at cubic order. Indeed, the remaining field-dependent
matrix elements connect only neighboring winding sectors. With
\begin{align}
&U_\pi = \bigoplus_{n=-N}^{N}(-1)^n I_4, \\
&U_\pi \mathbb K_{N,{\rm off}}(\phi,\mathbf B_\parallel)U_\pi^\dagger
  = \mathbb K_{N,{\rm off}}(\phi,-\mathbf B_\parallel),
\label{eq:lp_tensorial_off_parity}
\end{align}
and $[U_\pi,\mathbb S_N]=0$, the generalized temperature spectrum is even in the in-plane field. Magnetic time reversal implies $T_c(\phi,\mathbf B_\parallel)=T_c(-\phi,\mathbf B_\parallel)$. Additional zero-field winding couplings need not preserve this extra relation, so this result is specific to the retained functional. Conversely, a nonzero isolated loop amplitude is not sufficient: the observable response depends on the leading mode and the sum over all contributing paths.

For $\eta_{1E}=\eta_{2E}=0$, as in Table~\ref{tab:gl_params}, two further operations act in the same way. The sign reversal of the doublet, $U_E=\bigoplus_n\mathrm{diag}(1,1,-1,-1)$, maps $\mathbb K_N(\phi,\mathbf B_\parallel)$ to $\mathbb K_N(\phi,-\mathbf B_\parallel)$ when the helical couplings vanish, and $U_\pi U_E$ does so when the non-derivative couplings vanish. Each group of field-linear couplings is required to all orders in $B$ in this case; zero-field mixing lifts the last two constraints. Without a symmetry reason, $c_3$ can also vanish: increasing $|\xi''_E|$ by $18\%$ gives $c_3=0$, and $O_{1/4}$ starts as $B^5$ there but at $B=0.75$ still reaches about $60\%$ of its value for Table~\ref{tab:gl_params}.

Under the homogeneous ideal-ring assumptions used here, the parity of Eq.~\eqref{eq:lp_tensorial_off_parity} holds for any single-component order parameter, whatever its representation and spin content. Time reversal makes the zero-field coefficients even in momentum, $\alpha(\mathbf q)=\alpha(-\mathbf q)$, so zero-field terms change the winding number by multiples of six and terms odd in the field by odd amounts. As a result $U_\pi$ maps $\mathbf B_\parallel$ to $-\mathbf B_\parallel$ to all orders, and the LP curve is reciprocal. Accordingly, the $A_1$, $A_2$, or $E$ channel alone, or $A_1$ and $A_2$ without coupling to $E$, gives no flux-odd response within numerical accuracy. Boundary or spatial structure beyond these assumptions, for example trigonal anisotropy of the ring edges, can lift this restriction.

The LP signatures also do not single out the parent representation: interchanging the bare transition temperature of the $A_2$ channel with that of the $A_1$ or the $E$ channel, at fixed couplings, gives a flux-odd response of comparable or larger size and for these couplings the $A_2$ parent yields the weakest response. The substitution $\psi_1\leftrightarrow\psi_2$, $\boldsymbol\psi_E\to\hat{\mathbf z}\times\boldsymbol\psi_E$ maps each bulk invariant onto an invariant of the same type, because $A_2\otimes A_1=A_2$, $A_2\otimes A_2=A_1$, and $A_2\otimes E=E$. With the edge prescription of Appendix~\ref{app:quadratic_invariants} the mixed scalar invariant~\eqref{eq:app_F12L} is not closed under this substitution. For $L_B=(\mathbf B_\parallel\cdot\mathbf t)D_s$, the ring inner product gives $L_B^\dagger=-L_B+(\mathbf B_\parallel\cdot\mathbf n)/R$, so the substitution generates an additional curvature term proportional to $(i\xi_{12}/R)(\mathbf B_\parallel\cdot\mathbf n)(\psi_1^*\psi_2-\psi_2^*\psi_1)$. Only when this symmetry-allowed term is included does the $A_2$-parent model map onto an $A_1$-parent model with identical $T_c(\phi,\mathbf B_\parallel)$.

To calculate the contribution proportional to $B^3$, we keep the full dependence of the coupling matrix elements on the winding wave numbers $k_n=(n+\phi)/R$ and define the Hermitian
temperature operator
\begin{equation}
\begin{split}
\mathcal H(\phi,B,\theta_B)
&=-\mathbb S_N^{-1/2}
\mathbb K_N^{(0)}(\phi,B,\theta_B)
\mathbb S_N^{-1/2}\\
&=\mathcal H_0(\phi)+B\,\mathcal V(\phi,\theta_B).
\end{split}
\label{eq:lp_temperature_operator}
\end{equation}
Since $\mathbb S_N$ is positive definite, this transformation preserves the generalized eigenvalues of Eq.~\eqref{eq:gen_eig}, where the largest one is $T_c$. Let $\tau_j$ and $|u_j\rangle$ denote the eigenvalues and normalized eigenvectors of $\mathcal H_0$, with $\tau_0$ the nondegenerate leading eigenvalue, and define $\mathcal V_{ij}=\langle u_i|\mathcal V|u_j\rangle$.
The zero-field operator $\mathcal H_0$ is diagonal in winding, whereas $\mathcal V$ changes winding by $\pm1$ or $\pm2$, so
$\mathcal V_{00}=0$. The third-order correction is given by
\begin{equation}
 \delta\tau_0^{(3)}
 =B^3\operatorname{Re}
 \sum_{j,k\ne0}
 \frac{\mathcal V_{0j}\mathcal V_{jk}\mathcal V_{k0}}
 {(\tau_0-\tau_j)(\tau_0-\tau_k)}.
 \label{eq:lp_cubic_eigenvalue}
\end{equation}
All couplings and eigenvalue separations preserve their actual flux and radius dependence. The flux-odd contribution follows by antisymmetrizing this correction between $\phi$ and $-\phi$.
Each nonzero product represents a closed three-step winding process. The three-step sequences in Eq.~\eqref{eq:closed_winding_sequences} generate the cubic trigonal factor. The full sum determines its coefficient and permits cancellations between different paths.
The pair-breaking term in Eq.~\eqref{eq:pair_breaking} adds $B^2\mathcal W$ to the temperature operator in Eq.~\eqref{eq:lp_temperature_operator} so that
\begin{equation}
\mathcal H(\varphi,B,\theta_B) =\mathcal H_0(\varphi)+B\,\mathcal V(\varphi,\theta_B) +B^2\mathcal W.
\end{equation}
The field-independent operator $\mathcal W$ is
\begin{equation}
\mathcal W = -\mathbb S_N^{-1/2} \Bigl[ \mathbf 1_{2N+1}\otimes \operatorname{diag}(\beta_1,\beta_2,\beta_E,\beta_E) \Bigr] \mathbb S_N^{-1/2}.
\end{equation}
The minus sign follows from the definition of the temperature
operator: positive pair-breaking coefficients increase the
quadratic GL energy and lower the transition temperature.
In the orthonormal eigenbasis of $\mathcal H_0$, we define
\begin{equation}
\mathcal W_{ij} = \langle u_i|\mathcal W|u_j\rangle.
\end{equation}
The operator $\mathcal W$ preserves the winding number,
although it need not be diagonal in the eigenmode index.
Since $\mathcal H_0$ is diagonal in winding, its eigenvectors
can be chosen with definite winding number. The operator
$\mathcal V$ connects sectors differing by $\Delta n=\pm1,\pm2$, whereas $\mathcal W$ connects only states in the same sector. Consequently, their mixed third-order contribution vanishes:
\begin{equation}
\sum_{j\neq0}
\frac{
\mathcal V_{0j}\mathcal W_{j0}
+\mathcal W_{0j}\mathcal V_{j0}
}{
\tau_0-\tau_j
}
=0.
 \label{eq:app_pair_breaking_selection}
\end{equation}
Such terms leave the cubic coefficient unchanged. Since the flux-odd part of $T_c$ is odd in $B$, they first modify it at order $B^5$. General symmetry-allowed field-quadratic invariants can mix winding sectors and change the cubic coefficient, but they are not included here.

The expansion applies when field-induced mixing is small compared with the relevant zero-field eigenvalue separations. Near a mode reorganization, the full diagonalization is required. Both the coupling matrix elements and the eigenvalue
differences in the denominators depend on the ring radius. Therefore, this expression does not guarantee that $A_{\rm nr}$ has a maximum at a finite radius. The distinct radius dependences of $A_{\rm nr}$ and $E$ are illustrated in Fig.~\ref{fig:radius_scan}.

For the Table~\ref{tab:gl_params} coefficients, $R=1.25$, and $\theta_B=\pi/6$, an independent implementation of the same Cartesian projection at the Fig.~\ref{fig:cubic_scaling} resolution gives $A_{\rm nr}\propto B^p$ with $p=3.00$ from fields between $B=0.01$ and $0.04$. At these fields the true maximum lies within $10^{-3}$ of half-integer flux, where the fixed-flux expansion is not uniform. Locally refined maxima give $p=2.99$. Evaluating Eq.~\eqref{eq:lp_cubic_eigenvalue} at the fixed reference flux
$\phi_0=1/4$ gives, independently of a field fit,
\begin{equation}
 O_{1/4}(B)=-4.84\times10^{-4}B^3+\mathcal O(B^5).
 \label{eq:lp_fixed_flux_cubic_check}
\end{equation}
The fixed-flux coefficient provides a separate check that does not require maximizing over a discrete flux grid. The cubic asymptote should be distinguished from the finite-field amplitude and polarity changes in Fig.~\ref{fig:Anr_signed_maps}.

\bibliography{refs}

\end{document}